\documentclass[pra,aps,twocolumn,preprintnumbers,showpacs,superscriptaddress]{revtex4}
\usepackage{latexsym,epsfig,amssymb,amsfonts,amsmath,graphicx,bbm,mathptm,comment,hyperref}
\pdfoutput=1

\usepackage{color}
\definecolor{dred}{rgb}{.8,0.2,.2}
\definecolor{dyellow}{rgb}{.7,.7,.0}
\definecolor{ddred}{rgb}{.4,.0,.0}
\definecolor{dblue}{rgb}{.2,.2,.8}

\newcommand{\RR}{\mathbf{R}}

\newcommand{\dd}{\; \mathrm{d}}

\newcommand{\ket}[1]{ |  #1 \rangle}

\newcommand{\bra}[1]{ \langle #1  |}

\newcommand{\ee}{\textrm{e}}

\newcommand{\bb}{\mathbf{b}}
\newcommand{\GG}{\mathbf{G}}
\newcommand{\kk}{\mathbf{k}}
\newcommand{\rr}{\mathbf{r}}

\newcommand{\AAA}{\mathbf{A}}
\newcommand{\BB}{\mathbf{B}}
\newcommand{\0}{\mathbf{0}}

\newcommand{\MM}{\texttt{M}}

\newcommand{\eqr}[1]{Eq.~(\ref{#1})}
\newcommand{\fir}[1]{Fig.~\ref{#1}}
\newcommand{\secr}[1]{Sec.~\ref{#1}}
\newcommand{\apr}[1]{Appendix~\ref{#1}}

\allowdisplaybreaks

\begin{document}

\title{Ab initio derivation of Hubbard models for cold atoms in optical lattices}

\author{R. Walters}
\affiliation{Clarendon Laboratory, University of Oxford, Parks Road, Oxford OX1 3PU, United Kingdom}
\author{G. Cotugno}
\email{g.cotugno1@physics.ox.ac.uk}
\affiliation{Max Planck Research Department for Structural Dynamics, University of Hamburg, CFEL, Hamburg, Germany}
\affiliation{Clarendon Laboratory, University of Oxford, Parks Road, Oxford OX1 3PU, United Kingdom}
\author{T. H. Johnson}
\affiliation{Clarendon Laboratory, University of Oxford, Parks Road, Oxford OX1 3PU, United Kingdom}
\author{S. R. Clark}
\affiliation{Centre for Quantum Technologies, National University of Singapore, 3 Science Drive 2, 117543, Singapore}
\affiliation{Clarendon Laboratory, University of Oxford, Parks Road, Oxford OX1 3PU, United Kingdom}
\affiliation{Keble College, University of Oxford, Parks Road, Oxford OX1 3PG, United Kingdom}
\author{D. Jaksch}
\affiliation{Clarendon Laboratory, University of Oxford, Parks Road, Oxford OX1 3PU, United Kingdom}
\affiliation{Centre for Quantum Technologies, National University of Singapore, 3 Science Drive 2, 117543, Singapore}
\affiliation{Keble College, University of Oxford, Parks Road, Oxford OX1 3PG, United Kingdom}

\date{\today}

\begin{abstract}
We derive {\em ab initio} local Hubbard models for several optical lattice potentials of current interest, including the honeycomb and Kagom\'{e} lattices, verifying their accuracy on each occasion by comparing the interpolated band structures against the originals. To achieve this, we calculate the maximally-localized generalized Wannier basis by implementing the steepest-descent algorithm of Marzari and Vanderbilt [N. Marzari and D. Vanderbilt, Phys. Rev. B {\bf 56}, 12847 (1997)] directly in one and two dimensions. To avoid local minima we develop an initialization procedure that is both robust and requires no prior knowledge of the optimal Wannier basis. The MATLAB code that implements our full procedure is freely available online at \url{http://ccpforge.cse.rl.ac.uk/gf/project/mlgws/}.
\end{abstract}


\maketitle
Atoms loaded into periodic optical potentials~\cite{bloch2008,lewenstein2012} can be sufficiently cold to only occupy a small number of lowest energy Bloch bands. The interaction between two atoms occupying the same potential well can be large so that they form a paradigm test-bed for studying the physics of strongly correlated quantum lattice models. To derive accurate microscopic models it is desirable to express the state of the atoms in terms of a basis of highly-localized single-particle states, given by some unitary transformation of the Bloch states forming the lowest energy bands. The reasons for this are two-fold. First, the occupations of localized basis states are measurable through high-resolution imaging~\cite{bakr2009}. Second, the Hamiltonian rewritten in terms of localized basis states is typically a Hubbard model dominated by a few local terms. Together these two points justify the simulation of local Hubbard models, used to describe many phenomena in condensed matter, using cold atoms in optical lattices~\cite{lewenstein2012}. In this article, we develop a procedure to systematically find a set of highly-localized basis states and thereby derive {\em ab initio} the parameters of a Hubbard model realized using cold atoms and an optical lattice.

Only in simple cases, e.g., a lattice potential that is orthogonal~\cite{jaksch1998} or leads to an isolated lowest Bloch band~\cite{blakie2004}, have the parameters of Hubbard models realized by cold atoms in optical lattices been derived using a basis of localized single-particle states. The single-particle states used are Fourier transforms of the Bloch states, called Wannier states~\cite{wannier1937}. For more complicated optical-lattice potentials, Hubbard parameters have been estimated rather than derived from first principles: on-site interaction Hubbard parameters have been estimated by using Gaussians centered at lattice minima as approximations to the single-particle states, and nearest-neighbor hopping parameters found by fitting a tight-binding form to the energy structure of the bands, without a rigorous justification of the tight-binding assumption (see e.g. Refs.~\cite{bloch2008, damski2005, santos2004}). 
The approach we take here improves upon such calculations in two ways. We use a class of single-particle states that generalize the Wannier states and can thus be more localized. Also, our procedure calculates Hubbard parameters from first principles, without approximation, and provides a quantitative justification of neglected terms. The necessity of such improvements has recently been noted in the literature~\cite{mering2011}.

Our procedure is an adaptation of several others already in use in solid-state physics. Specifically, we take as our starting point an algorithm developed by Marzari and Vanderbilt~\cite{marzari1997}. They consider a basis of generalized Wannier states; Fourier transforms of inter-band mixtures of Bloch states. Choosing some initial basis, a steepest-descent minimization algorithm is used to iteratively generate another set of generalized Wannier states with a smaller spatial spread. The desired end-point of these iterations is the basis corresponding to the global minimum of the spread, the so-called maximally-localized generalized Wannier states (see Ref.~\cite{marzariRev} and references within for a review on the topic). Once this optimal basis is found, the parameters of the corresponding Hubbard model are easily calculated.

The currently available software packages \cite{wannierorg} that implement the steepest-descent minimization algorithm operate in three dimensions. For use with optical lattice potentials, which are often effectively one or two-dimensional, we have implemented the algorithm directly in these lower dimensional spaces, as well as in three dimensions. We find that for the optical-lattice potentials considered here, our implementation in conjunction with commonly used initialization procedures (e.g. that described in~\cite{marzari1997}) typically fails to converge to the global minimum of the spread and instead becomes trapped in a local minimum; the maximally-localized generalized Wannier states are not obtained. 

Therefore, our algorithmic contribution is a new initialization procedure for the Marzari and Vanderbilt steepest-descent algorithm. Our initialization procedure has an additional benefit in that it requires no knowledge of the optimal Wannier states, e.g., their location or approximate form, and therefore requires no input beyond specifying the lattice potential. The initialization procedure is split into two parts, each minimizing the inter- and intra-band contributions to the spread of the generalized Wannier states, respectively. The former is a method for minimizing the spread in the case of a single band~\cite{marzari1997}. The latter relates to a procedure devised by Souza, Mazari and Vanderbilt to optimally disentangle a subset of bands from a group of degenerate bands~\cite{souza2001}. Our whole procedure, taking the lattice potential as input, and outputting the maximally localized Wannier states and Hubbard parameters, is combined into a single MATLAB routine. We have made this code freely available online~\cite{tnt}.

Note that while in the last stages of preparing this article we became aware of a very recent article~\cite{azpiroz2012} in which the authors use a different procedure to compute the maximally-localized generalized Wannier states and justify a local Hubbard model for bosons in the two-dimensional honeycomb potential.

The remainder of the article is organized as follows. In \secr{sec:problem} we discuss the derivation of Hubbard models for cold atoms in optical lattices, introducing generalized Wannier states as a basis for this derivation and outlining the problem of finding the states with minimum combined spread. Our approach for obtaining the maximally-localized basis is then described in \secr{sec:methods}. We include an outline of Marzari and Vanderbilt's steepest-descent algorithm, discuss the steps of our initialization procedure and then summarize how we combine these elements. In \secr{sec:results} we derive Hubbard models for bosons in several optical-lattice potentials, first in one dimension then in two, verifying the accuracy of our calculations on each occasion. Finally, we conclude in \secr{sec:conclusions} before presenting computational details in the appendices.

\section{Objective}\label{sec:problem}

\subsection{Hubbard models for atoms in optical lattices}
To begin, we outline the typical approach to deriving Hubbard models for ultracold atoms with mass $\mu$ in an optical lattice. For simplicity we assume the atoms to be spinless bosons; extensions to fermionic atoms, multi-component gases including Bose-Fermi mixtures with different lattice potentials, atom-molecular interactions and finite-range interactions are straightforward \cite{lewenstein2012}.

Standing waves of laser light, tuned out of resonance, exert a spatially-periodic AC Stark shift on the ground internal state of the bosons. For sufficiently low atom energies and densities $\rho$ the interactions between the atoms are well-approximated by a contact interaction of strength $g$. The effective Hamiltonian is then of the form \cite{jaksch1998}
\begin{equation*}
\hat{H} = \int \dd \mathbf{r} \; \hat{\Psi}^{\dagger}(\mathbf{r}) \hat{h} \hat{\Psi}(\mathbf{r}) + \frac{g}{2} \int \dd \mathbf{r} \;  \hat{\Psi}^{\dagger}(\mathbf{r}) \hat{\Psi}^{\dagger}(\mathbf{r}) \hat{\Psi}(\mathbf{r}) \hat{\Psi}(\mathbf{r}) .
\end{equation*}
Here $\hat{\Psi}$ annihilates a boson of mass $\mu$ and the single-particle Hamiltonian is $\hat{h} = - \hbar^2 \nabla^2 /2\mu + V (\mathbf{r})$, where $V (\mathbf{r})$ is the lattice potential induced by the AC Stark shift.

We expand the field operators in terms of a complete basis of orthonormal mode functions $w^{n}_{\RR} (\mathbf{r})$, corresponding to single-particle states
\begin{align*}
\ket{ \RR n} = \int \dd \mathbf{r} \; w^{n}_{\RR} (\mathbf{r}) | \mathbf{r} \rangle ,
\end{align*}
obeying the translational equivalence
\begin{equation}
\label{eq:translation}
w^{n}_{\RR}(\mathbf{r}) = w^{n}_{\RR'} (\mathbf{r} + \mathbf{R}' - \mathbf{R}) .
\end{equation}
Here $\RR$ is a direct lattice vector for which $V (\mathbf{r}+\mathbf{R}) = V(\mathbf{r})$ is satisfied, and which indicates the lattice site where $w^{n}_{\RR}(\mathbf{r})$ is localized, relative to some origin. The integer $n$ is commonly called the band number, although as we shall see shortly it will index modes which may comprise of mixtures of several bands. An atom occupying the mode $w^{n}_{\RR}(\mathbf{r})$ is often said to be in the $n$-th excited state or mode of lattice site $\RR$.

The expansion thus takes the form
\begin{eqnarray}
\hat{\Psi} (\mathbf{r}) &=& \sum_\RR  \sum_{n} w^{n}_{\RR}  (\mathbf{r}) \hat{b}^{n}_{\RR}  \nonumber,
\end{eqnarray}
where $\hat{b}^{n}_{\RR}$ annihilates a boson in mode $w^{n}_{\RR}(\mathbf{r})$, such that the Hamiltonian $\hat{H}$ may be re-expressed as
\begin{eqnarray*}
\hat{H} &=&  - \sum_{mn} \sum_{\RR \RR'} t^{mn}_{\RR \RR'} \hat{b}^{m \dagger}_{\RR} \hat{b}^{n}_{\RR'} \nonumber \\
&&+ \frac{1}{2} \sum_{mnop} \sum_{\RR \RR' \RR'' \RR'''} U^{mnop}_{\RR \RR' \RR'' \RR'''} \hat{b}^{m \dagger}_{\RR} \hat{b}^{n \dagger}_{\RR'} \hat{b}^{o}_{\RR''} \hat{b}^{p}_{\RR'''} ,
\end{eqnarray*}
with hopping and interaction parameters
\begin{eqnarray}
t^{mn}_{\RR \RR'} &=&  - \int \dd \mathbf{r} \; w^{m \ast}_{\RR}(\mathbf{r})  \hat{h} w^{n}_{\RR'}(\mathbf{r}) \nonumber, \\
U^{mnop}_{\RR \RR' \RR'' \RR'''} &=& g \int \dd \mathbf{r} \; w^{m \ast}_{\RR}(\mathbf{r}) w^{n \ast}_{\RR'}(\mathbf{r}) w^{o}_{\RR''}(\mathbf{r}) w^{p}_{\RR'''}(\mathbf{r}) \nonumber.
\end{eqnarray}
Due to \eqr{eq:translation}, these parameters are invariant under a simultaneous translation in the direct lattice vectors that label them.

The Hamiltonian simplifies in two ways. First, for sufficiently small kinetic $E_{\rm kin}$ and interaction energies $E_{\rm int} \approx \rho g$, we can ignore all but some number $J$ of the bands. Second, $w^{n}_\RR (\mathbf{r})$ are chosen such that they are well localized, meaning that the $t^{mn}_{\RR \RR'} $ and $U^{mnop}_{\RR \RR' \RR'' \RR'''}$ corresponding to hopping or interaction between distant states are negligible. This leaves the Hubbard model
\begin{eqnarray}
\hat{H}_{\rm HM} &=&  - \sum_{mn = 1 }^J   \sum_{\langle \RR \RR' \rangle} t^{mn}_{\RR \RR'} \hat{b}^{m \dagger}_{\RR} \hat{b}^{n}_{\RR'} \nonumber \\
&&+ \frac{1}{2} \sum_{ mnop = 1}^J \sum_{\langle \RR \RR' \RR'' \RR''' \rangle} U^{mnop}_{\RR \RR' \RR'' \RR'''} \hat{b}^{m \dagger}_{\RR} \hat{b}^{n \dagger}_{\RR'} \hat{b}^{o}_{\RR''} \hat{b}^{p}_{\RR'''} \nonumber,
\end{eqnarray}
where the angular brackets indicate that the sum is restricted to local terms, e.g., same-site, nearest-neighbor, or next-nearest-neighbor etc. The range of the terms that need to be kept will depend on how local the $w^{n}_{\RR}(\mathbf{r})$ can be, which in turn is dependent on the form of the potential $V (\mathbf{r})$.

\subsection{Generalized Wannier states}
We now turn our attention to the choice of wavefunctions $w^{n}_{\RR}(\mathbf{r})$ used in the above procedure. A complete basis of orthonormal functions is provided by the Bloch states $\ket{\psi^{(\kk)}_{m}}$, corresponding to eigenfunctions of $\hat{h}$
\begin{align*}
\psi^{(\kk)}_{m} (\mathbf{r})=  \ee^{\mathrm{i} \kk \cdot \rr } u^{(\kk)}_{m} (\mathbf{r}),
\end{align*}
where $u^{(\kk)}_{m} (\mathbf{r})$ are cell-periodic functions~\cite{ashcroft}. The Bloch states of a given band $m$ are uniquely labeled by a wave-vector $\kk$ that runs over the first Brillouin zone of the reciprocal lattice. Any band $m$ with energies
\begin{align*}
E^{(\kk)}_{m} =  \bra{\psi^{(\kk)}_{m}} \hat{h} \ket{\psi^{(\kk)}_{m}} ,
\end{align*}
satisfying $E^{(\kk)}_{m} \gg E_{\rm kin}, \, E_{\rm int}$ for all $\kk$ will not contribute to the physics and may be ignored. For all optical lattice potentials we consider here it is possible to focus solely on a small number $J$ of the lowest-energy bands which may be degenerate amongst themselves but are separated in energy from the others.

To describe local interactions within this $J$-band subspace, a good choice of basis are states of the form
\begin{equation}\label{eq:wannier_generalized}
\ket{\RR n} = \frac{\Upsilon}{(2\pi)^{D}} \int_{\mathrm{BZ}} \dd \kk \; \ee^{-\mathrm{i} \kk \cdot \RR} \sum_{m=1}^J U^{(\kk)}_{m n} \ket{ \psi^{(\kk)}_{m} },
\end{equation}
where $\Upsilon$ is the volume of the primitive cell of the $D$-dimensional direct lattice, and $U^{(\kk)}$ is a unitary matrix that mixes the Bloch bands.
In the case that $U^{(\kk)}$ is diagonal, i.e., there is no band mixing, these states are exactly those first considered by Wannier~\cite{wannier1937}. Thus the states appearing in \eqr{eq:wannier_generalized} are commonly referred to as generalized Wannier states.

The separation in energy of the $J$ lowest bands from the others ensures that some states $\ket{\RR n}$ exist with mode functions $w^{n}_{\RR}(\mathbf{r})$ that are exponentially localized at lattice site $\RR$ in coordinate space~\cite{kohn1959, desCloizeaux1964,nenciu1983,brouder2007,panati2007}. This exponential localization occurs if and only if the Bloch superpositions
\begin{equation}\label{eq:bloch_superpositions}
\ket{\tilde{\psi}^{(\kk)}_{n}} = \sum_{m=1}^J U^{(\kk)}_{m n} \ket{\psi^{(\kk)}_{m}} ,
\end{equation}
are analytic (infinitely differentiable) in $\kk$ across the whole Brillouin zone~\cite{duffin1953}. This is a rigorous way of saying that only smoothed-out Bloch superpositions will lead to localization when Fourier transformed. When there are no degeneracies between bands can one simply use the phases of elements of a diagonal $U^{(\kk)}$ (representing the freedom in the phase of each $\ket{\psi^{(\kk)}_{m}}$) to ensure the smoothness of the Bloch states $\ket{\tilde{\psi}^{(\kk)}_{n}}$. Hence simple Wannier states provide an exponentially localized basis in such cases. However, this is no longer the case when degeneracies and crossings in the band structure lead to non-analytic $\ket{\psi^{(\kk)}_{m}}$. In this situation band mixing and therefore a non-diagonal $U^{(\kk)}$ are required to obtain smooth Bloch superpositions and an exponentially-localized basis. The `only if' case highlights the importance of the generalization of Wannier states to include non-diagonal $U^{(\kk)}$. Even when exponential localization is possible using simple Wannier states, generalized Wannier states may still significantly improve the localization. We will give examples of this in \secr{sec:results}.

\subsection{Maximally-localized generalized Wannier states}
Generalized Wannier states therefore have the potential to provide a well-localized basis for the derivation of a Hubbard model. However, generalized Wannier states are highly non-unique and so it remains to find and choose a single exponentially-localized basis.

Several criteria have been proposed as a means of selecting a specific basis of generalized Wannier states~\cite{stephan2000, andersen2000, marzari1997}. Here, following Ref.~\cite{marzari1997}, we seek the generalized Wannier states with a minimal combined spatial variance, henceforth called spread, defined as
\begin{eqnarray}\label{spread_total}
\Omega =& \sum_{n=1}^{J} \left[\langle \0 n |\hat{\mathbf{r}}^{2}| \0 n\rangle - \langle \0 n |\hat{\mathbf{r}}| \0 n\rangle^2 \right] \nonumber\\
=& \sum_{n=1}^{J}   \left[  \langle \rr^2 \rangle_n - \bar{\rr}_n^2 \right] = \sum_{n=1}^{J} \Omega^n .
\end{eqnarray}
The minimizing states are called the maximally-localized generalized Wannier states.

It is known that the maximally-localized generalized Wannier states are indeed exponentially localized~\cite{kohn1959, desCloizeaux1964, nenciu1983, he2001, panati2011}. They do not necessarily provide the optimal approximation to $\hat{H}$ when restricting the number of terms kept in $\hat{H}_{\rm HM}$ but can be expected to be close to optimal. Finding a set of generalized Wannier states which makes the Hubbard model approximation optimal is challenging. Hence minimizing the spread of the generalized Wannier functions is both an effective and practical choice with the added benefit of having a straightforward interpretation in terms of a particle occupying a specific lattice site. Note that it has been hypothesized that the maximally-localized generalized Wannier states always correspond to real functions, up to a global phase, when dealing with an isolated group of $J$ bands~\cite{zak1982, brouder2007, panati2011}. All of our calculations support this hypothesis.

\section{Method}\label{sec:methods}
To obtain maximally localized Wannier states we must first calculate the band structure $E^{(\kk)}_{m}$ and Bloch states $\ket{\psi^{(\kk)}_{m}}$. Such calculations are well-understood and we include our procedure here only for completeness and to introduce notation required later. Then we must find the gauge $U^{(\kk)}$ such that the resulting generalized Wannier states $\ket{\RR n}$, defined by \eqr{eq:wannier_generalized}, minimize the spread, defined by \eqr{spread_total}. This usually consists of several steps: initially, we choose the gauge $U^{(\kk)} = \mathbbm{1}_J$ to be the $J\times J$ identity matrix. Then the gauge is transformed iteratively $U^{(\kk)} \rightarrow U^{(\kk)} V^{(\kk)}$ according to a unitary $V^{(\kk)}$. These transformations accumulate until they converge to the desired gauge $U^{(\kk)}$, corresponding to the minimum spread. From this, the maximally-localized generalized Wannier states may be calculated together with the hopping and interaction parameters.

In the strategy devised by Mazari and Vanderbilt~\cite{marzari1997} an initial unitary $V^{(\kk)}$ is constructed via projections of $J$ localized trial orbitals onto the Bloch states. This transformation leads to a gauge $U^{(\kk)} = V^{(\kk)}$ corresponding to an analytic set of Bloch superpositions $\ket{\tilde{\psi}^{(\kk)}_{n}}$ and thus exponentially-localized generalized Wannier states $\ket{\RR n}$. Subsequently, other unitary transformations $V^{(\kk)}$ are iteratively applied as part of a steepest-descent algorithm, in the hope that the cumulative gauge $U^{(\kk)}$ converges towards the spread-minimizing gauge.

Unfortunately when applying this strategy to investigate common optical-lattice potentials, we found that the gauge corresponding to the maximally-localized generalized Wannier states is rarely obtained. Instead the spread often converges to some non-global minimum. Therefore we adopt a different initialization procedure, to precede the same steepest-descent algorithm. Contrastingly, we find that, for the optical-lattice potentials considered, our strategy converges quite consistently to the global minimum, and thus the maximally-localized generalized Wannier states are reliably obtained. While proving convergence to the global minimum is difficult, we test numerically both that the same solution is found when starting from two Bloch states differing by random permutations of the bands at each $\kk$, and that the final generalized Wannier states are real.

In this section we begin by outlining the band structure calculation, and the representation of quantities, such as the spread, in reciprocal space. We then briefly describe Mazari and Vanderbilt's steepest-descent algorithm, before outlining two methods we will use as our initialization procedure. To end the section, we describe our full procedure for calculating the maximally-localized generalized Wannier states.

\subsection{Band structure}\label{app:mesh_bz}
\begin{figure}
\begin{center}
\includegraphics[width=8.5cm]{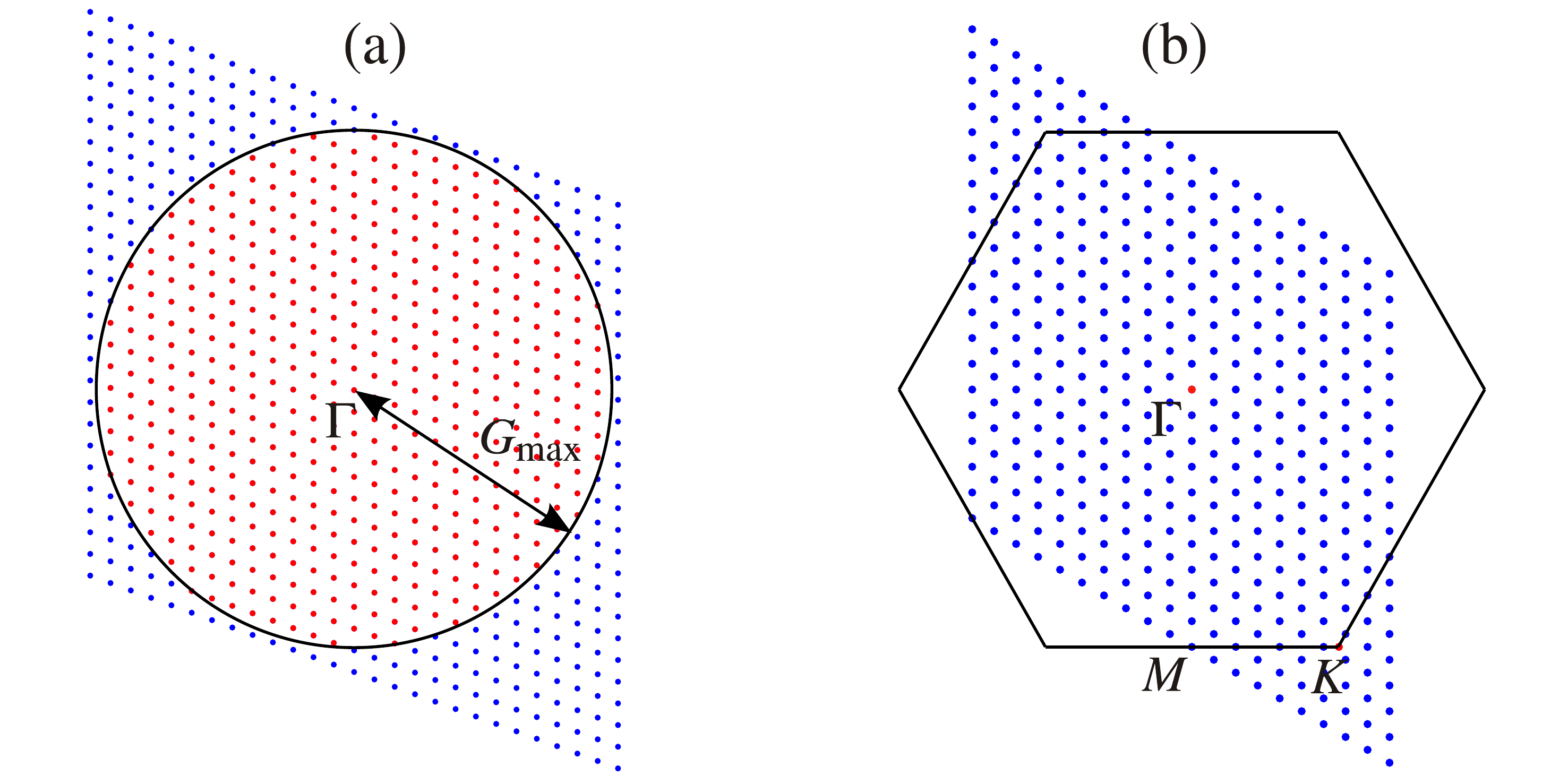}
\caption{(Color online) Representing the problem in $\kk$-space. (a) The reciprocal lattice points for a two-dimensional oblique lattice. The black circle has a radius of $G_\textrm{max}$ and is centered on $\Gamma$. Fourier components corresponding to reciprocal lattice points within the circle (red points) are included in the truncated basis set, while those outside (blue points) are not. (b) A mesh of wave-vectors $\kk$ for a hexagonal two-dimensional lattice used to interpolate values of functions of $\kk$ over the Brillouin zone, which is shown by the black hexagon. }
\label{Fig:k-mesh}
\end{center}
\end{figure}

We work with the Fourier space representation of the potential and cell-periodic functions
\begin{eqnarray}
v^{(\GG)} &=& \frac{1}{\sqrt{\Upsilon}}\int_{\mathrm{PC}} \dd \textbf{r} \; V (\textbf{r}) \ee^{-\mathrm{i}\textbf{G} \cdot\textbf{r}} , \nonumber \\
c^{(\kk, \GG)}_{m} &=& \frac{1}{\sqrt{\Upsilon}} \int_{\mathrm{PC}} \dd \textbf{r} \; u^{(\kk)}_{m} (\textbf{r}) \ee^{-\mathrm{i}\textbf{G} \cdot\textbf{r}} \nonumber,
\end{eqnarray}
where the integral is over a primitive cell of the direct lattice. In this representation the single-particle Schr\"{o}dinger equation $\hat{h} \ket{\psi^{(\kk)}_{m}} = E^{(\kk)}_{m} \ket{\psi^{(\kk)}_{m}}$ may be written as
\begin{equation}
\frac{1}{2M}(\textbf{G}+\textbf{k})^2 c^{(\kk, \GG)}_{m} + \frac{1}{\sqrt{\Upsilon}}\sum_{\GG'} v^{(\GG-\GG')} c^{(\kk, \GG')}_{m} = E^{(\kk)}_{m} c^{(\kk, \GG)}_{m}.
\label{coeff_eqn}
\end{equation}
The full band structure is obtained by solving this equation for all wave-vectors $\textbf{k}$ in the Brillouin zone and all reciprocal lattice vectors $\GG$~\cite{ashcroft}.

To make this calculation tractable on a computer, we firstly truncate the Fourier expansions to include some finite number $\texttt{N}$ of terms, corresponding to reciprocal lattice vectors $\GG$ with magnitudes $|\GG|$ less than $G_\textrm{max}$, as shown in \fir{Fig:k-mesh}(a). Then we only solve \eqr{coeff_eqn} for a $D$-dimensional uniform discrete mesh of $\texttt{M}^D$ wave-vectors $\textbf{k} = \textbf{G}/ \MM$ contained within some primitive cell of the reciprocal lattice, and interpolate between these wave-vectors. As shown in \fir{Fig:k-mesh}(b), this primitive cell need not be the first Brillouin zone since the corresponding Bloch states are invariant when translated by a reciprocal lattice vector into the Brillouin zone. For each $\kk$ in the mesh, solving the set of Eqs.~(\ref{coeff_eqn}) then reduces to an eigenvalue problem. The justification of the truncation and mesh discretization, as well as the values of $G_\textrm{max}$ and $\texttt{M}$ we use are discussed in \apr{sec:discretization}.

Note that various symmetries guarantee certain properties of the coefficients $c^{(\textbf{k},\GG)}_{m}$~\cite{Martin2004}. For a given $m$ and $\kk$, time-reversal symmetry implies that we may choose $c^{(-\kk,-\GG)}_{m} = c^{(\kk,\GG)\ast}_{m}$. The addition of inversion symmetry allows us to set $c^{(\kk,\GG)}_{m}$ and $v^{(\GG)}$ as real up to a common phase factor and ensures $v^{(\GG)} = v^{(-\GG)}$. This implies that, when both symmetries are present, we may both reduce our mesh of reciprocal lattice vectors by nearly half, as the coefficients for $-\GG$ may be inferred from those for $\GG$, and restrict all quantities in the eigenvalue equation (\ref{coeff_eqn}) to be real, thus speeding up computations for each $\kk$.

\subsection{Contributions to the spread}
Following Mazari and Vanderbilt, the spread $\Omega=\Omega_{I}+\tilde{\Omega}$ can be conveniently decomposed into two positive definite parts, $\Omega_{I}$ and $\tilde{\Omega}$. The latter depends on the choice of gauge $U^{(\kk)}$ appearing in \eqr{eq:wannier_generalized}, while the former does not. The gauge-independent part $\Omega_{I}$ depends only on the smoothness in $\kk$-space of the underlying manifold of Bloch states, while the gauge-dependent part $\tilde{\Omega}$ depends on the additional smoothing achieved by applying phases to and mixing the Bloch states. In preparation for what follows, it is useful to further decompose $\Omega_I = \Omega_{I,D}+\Omega_{I,OD}$ and $\tilde{\Omega} = \Omega_{D}+\Omega_{OD}$ into band-diagonal and band-off-diagonal terms. For the diagonal terms $\Omega_{I,D} = \sum_{n} \Omega_{I,D}^{n}$ and $\Omega_{D} = \sum_{n} \Omega_{D}^{n}$ it makes sense to break them down into positive-definite contributions from each band. The decomposition is expressed neatly as
\begin{eqnarray}
\Omega= \overbrace{ \underbrace{\sum_{n} \Omega_{I,D}^{n}}_{\Omega_{I,D}} + \Omega_{I,OD} }^{\Omega_I} + \overbrace{\underbrace{\sum_{n} \Omega_{D}^{n}}_{\Omega_{D}} + \Omega_{OD}  }^{\tilde{\Omega}} \nonumber .
\end{eqnarray}
Our minimization method will of course leave $\Omega_I$ invariant, while minimizing $\tilde{\Omega}$. The Mazari and Vanderbilt steepest-descent algorithm iteratively minimizes $\tilde{\Omega}$ directly, while our initialization procedure is divided into two stages, one which reduces $\Omega_{OD}$ and another which minimizes $\Omega_{D}$.

In terms of generalized Wannier states, the contributions to the spread are written
\begin{subequations}\label{eq:spread_realspace}
\begin{eqnarray}
\Omega_{I,D}^{n} & = & \langle \0 n |\hat{\mathbf{r}}^2| \0 n \rangle - \sum_{\RR} |\langle \0 n |\hat{\mathbf{r}}| \RR n \rangle |^{2} ,  \\
\Omega_{I, OD} & = &  - \sum_n \sum_{m\neq n}\sum_\RR |\langle \0 m |\hat{\mathbf{r}}|\RR n \rangle |^{2} , \\
\Omega_{D}^{n} & = & \sum_{\RR \neq \0}  |\langle \0 n | \hat{\mathbf{r}}|\RR n \rangle|^{2} , \\
\Omega_{OD} & = & \sum_n \sum_{m\neq n} \sum_\RR |\langle \0 m | \hat{\mathbf{r}}| \RR n \rangle|^{2} .
\end{eqnarray}
\end{subequations}

Again, for computational tractability, we move to the truncated Fourier representation with a discretized mesh. In this, all integrals over the Brillouin zone are replaced by summations over the mesh,
\begin{displaymath}
\frac{\Upsilon}{(2\pi)^D}\int_{\mathrm{BZ}} \dd \textbf{k} \; \rightarrow\frac{1}{\MM^D}\sum_\textbf{k},
\end{displaymath}
and gradients represented by finite differences (the gradients in reciprocal space arise from moments in position space). We use the finite-difference expressions recommended by Marzari and Vanderbilt~\cite{marzari1997}, which have the property of transforming correctly under translations of the generalized Wannier states by a direct lattice vector. In this way, contributions to the spread are re-expressed as
\begin{subequations}\label{eq:spread_kspace}
\begin{eqnarray}
\Omega_{I,D}^{n} & = & \frac{1}{\MM^{D}} \sum_{\mathbf{k,b}} \omega_{\mathbf{b}} \Bigg( 1 - |M^{(\mathbf{k,b})}_{nn}|^{2}  \Bigg), \\
\Omega_{I, OD} & = & - \frac{1}{\MM^{D}} \sum_{\mathbf{k,b}} \omega_{\mathbf{b}}  \sum_n \sum_{m\neq n} |M^{(\mathbf{k,b})}_{mn}|^{2},  \\
\Omega_{D}^{n} & = &  - \frac{1}{\MM^{D}} \sum_{\mathbf{k,b}} \omega_{\mathbf{b}} \left( \text{Im}[\ln M^{(\mathbf{k,b})}_{nn}] + \mathbf{b}\cdot\overline{\mathbf{r}}_n \right)^2, \label{eq:spread_D_kspace} \\
\Omega_{OD} & = & \frac{1}{\MM^{D}} \sum_{\mathbf{k,b}} \omega_{\mathbf{b}} \sum_n \sum_{m\neq n} |M^{(\mathbf{k,b})}_{mn}|^{2}.
\end{eqnarray}
\end{subequations}
Here the vectors $\mathbf{b}$ connect each wave-vector $\mathbf{k}$ to its nearest-neighbors, $\omega_{\mathbf{b}}$ are factors that depend on the geometry of the mesh~\cite{mostofi2008}, and
\begin{eqnarray}
\overline{\mathbf{r}}_n & = & -\frac{1}{\MM^{D}} \sum_{\mathbf{k,b}} \omega_{\mathbf{b}} \mathbf{b} \text{Im}[\ln M^{(\mathbf{k,b})}_{nn}] \nonumber.
\end{eqnarray}

It is clear then that all the information about the spread is contained in the matrix elements
\begin{displaymath}
M^{(\mathbf{k,b})}_{mn} = \sum_{op} U^{(\kk)\ast}_{pm} \sum_\GG c^{(\textbf{k} , \GG )\ast}_{p}  c^{(\textbf{k} + \bb, \GG)}_{o}     U^{(\kk)}_{on},
\end{displaymath}
which are the truncated Fourier representation of the overlap $\langle \tilde{u}^{(\kk)}_{m} | \tilde{u}^{(\kk + \bb)}_{n} \rangle$, where similarly to \eqr{eq:bloch_superpositions},
\begin{equation}
\label{eq:periodic_superpositions}
\ket{\tilde{u}^{(\kk)}_{n}} = \sum_{m=1}^J U^{(\kk)}_{m n} \ket{u^{(\kk)}_{m}} ,
\end{equation}
with $\ket{u^{(\kk)}_{m}}$ the state associated with periodic function ${u}^{(\kk)}_{m} (\rr)$. These elements are initialized to $M^{(\mathbf{k,b})}_{mn}= \sum_\GG c^{(\textbf{k} , \GG )\ast}_{m}  c^{(\textbf{k} + \bb, \GG)}_{n}  $ when $U^{(\kk)} = \mathbbm{1}_J$. Then under a gauge transformation $U^{(\kk)} \rightarrow  U^{(\kk)} V^{(\kk)}$ they undergo the computationally simple transformation $M^{(\mathbf{k,b})} \rightarrow V^{(\kk) \dagger} M^{(\mathbf{k,b})} V^{(\kk + \bb)}$.

\subsection{Minimizing total spread}\label{sec:steepest_descent}
The gradient $\Gamma^{(\mathbf{k})} = d\Omega/dW^{(\mathbf{k})}$, embodying the change in spread due to a gauge transformation $V^{(\mathbf{k})} =  \ee^{dW^{(\mathbf{k})}}$, with $dW^{(\mathbf{k})}$ an infinitesimal anti-Hermitian matrix, can be efficiently calculated from the matrices $M^{(\mathbf{k,b})}$ (see Ref.~\cite{marzari1997} for details). The steepest-descent approach, as used in Refs.~\cite{marzari1997,souza2001}, then implements the gauge transformation $V^{(\mathbf{k})} =  \ee^{dW^{(\mathbf{k})}}$ with $dW^{(\mathbf{k})} = - \epsilon \Gamma^{(\kk)}$ and $\epsilon$ a small positive number. These steps are repeated until convergence is achieved.

To a large extent, the steepest-descent algorithm is only as good as its initialization procedure, since starting from an arbitrary set of generalized Wannier states, the algorithm is likely to drive the set towards one of the many local minima in the spread, rather than the global minimum. We do not discuss here the commonly used projection-based initialization procedure (see Ref.~\cite{marzariRev} for information on this) that we found to struggle for optical-lattice potentials. Instead we now discuss two other approaches for reducing the spread, which together will form the initialization procedure we use successfully for optical-lattice potentials.

\subsection{Reducing inter-band spread} \label{sec:disentangling}
We break down the task of finding the maximally-localized Wannier states into two stages. The first stage is to mix the bands to create a new set of pseudo-bands from which a maximally-localized ordinary Wannier states calculation is optimal (i.e.\ leading to the smallest possible spread). The second stage is to calculate the maximally-localized ordinary Wannier states using these pre-mixed bands as a starting point. The first stage corresponds to minimizing the off-diagonal term $\Omega_{OD}$, and the second to minimizing the diagonal term $\Omega_{D}$. Our initialization procedure is split accordingly: first we reduce (but not necessarily minimize) $\Omega_{OD}$, as described in this subsection; second we minimize $\Omega_{D}$, as described in the next subsection.

Our first goal is then to reduce the band-off-diagonal term $\Omega_{OD}$, which is equivalent to reducing $\Omega_{I, D}$. This equivalence is clear from the interpretation above, that reducing $\Omega_{OD}$ corresponds to optimizing the bands from which to perform a maximally-localized ordinary Wannier states calculation. Mathematically, it follows from observing that the band-off-diagonal parts of the gauge-invariant spread $\Omega_{I}$ and the gauge-dependent spread $\tilde{\Omega}$ are the negative of each other (see Eqs.~(\ref{eq:spread_realspace}b), (\ref{eq:spread_realspace}d) and (\ref{eq:spread_kspace}b), (\ref{eq:spread_kspace}d)): reducing $\Omega_{OD}$ is achieved by increasing the band-off-diagonal part $\Omega_{I, OD}$ of the gauge-invariant spread or, equivalently, reducing its diagonal part $\Omega_{I, D}$.

To reduce $\Omega_{I, D}$, we use a method devised by Souza, Mazari and Vanderbilt~\cite{souza2001}. For $K$ degenerate bands, their minimizes the contributions to $\Omega_{I}$ from a subset $K' <K$ bands obtained through a unitary mixing of these bands. The aim of this approach is then to construct the $K'$ bands with the smoothest $\kk$-space such that they provide the optimal set of $K'$ bands from which to construct localized generalized Wannier states (optimal in the sense of having the smallest possible gauge-invariant contribution to the
spread).

We use their approach to reduce $\Omega_{I, D}$ in the following way: First, we use the Souza {\em et al.}\ method to minimize $\Omega_{I, D}^{1}$ and therefore construct, from the $K = J$ bands, a single ($K' = 1$) band whose smoothness in $\kk$-space is optimum for constructing a localized Wannier state. Then, keeping this band fixed, we use the Souza {\em et al.}\ method again to minimize $\Omega_{I, D}^{2}$ and construct from the $K = J-1$ remaining bands a single ($K'=1$) band that is optimum for constructing a localized Wannier state. This is repeated in a similar fashion to obtain a third, fourth, etc. band until finally we use the Souza {\em et al.}\ method to minimize $\Omega_{I, D}^{J-1}$ and construct an optimized $(J-1)$-th band out of the two remaining bands, with all lower bands fixed. Our approach therefore consists of $J-1$ applications of the Souza {\em et al.}\ method, in each case optimally extracting a single ($K' = 1$) band from $K= J, J-1, \dots, 2$ others.

Note that this does not necessarily minimize $\Omega_{I, D}$ and therefore $\Omega_{OD}$, but we find that following this procedure $\Omega_{OD}$ is very small. Details of the Souza {\em et al.}\ method and our use of it can be found in Ref.~\cite{souza2001} and \apr{app:our_algorithm}, respectively. Here we simply note that the Souza {\em et al.}\ method proceeds via several iterations, each of which applies a transformation $V^{(\kk)}$ over all $\kk$-space that would have minimized the contribution to the spread from any given point in $\kk$-space had there been no transformation applied at the other points in $\kk$-space. The desired gauge must be left unchanged by such an iteration and thus it is a possible point of convergence. To protect against false convergences, we initialize the whole procedure above by applying a transformation $V^{(\mathbf{k})}$, where at each $\mathbf{k}$ we take a $J\times J$ identity matrix and randomly permute its rows. We find that in practice, following this initialization, the desired gauge is nearly always obtained.

\subsection{Reducing intra-band spread} \label{sec:isolated_bands}
We next present a method that reduces the intra-band contribution $\Omega_{D}$ to the spread, while leaving the inter-band contribution $\Omega_{OD}$ invariant. To ensure this invariance, in this section we restrict ourselves to gauge transformations $V^{(\kk)}$ that are diagonal, i.e., while we allow changes to the phases of the Bloch superpositions $\ket{\tilde{\psi}^{(\kk)}_{n}}$, we do not allow any transformations that further mix the bands. Hence the task splits into $J$ independent parts, each to reduce $\Omega^{n}_{D}$ by applying phases to the Bloch superpositions $\ket{\tilde{\psi}^{(\kk)}_{n}}$. One may interpret this as constructing the maximally-localized ordinary Wannier states from a set of bands comprising the mixed Bloch states $\ket{\tilde{\psi}^{(\kk)}_{n}}$.

Such single-band tasks are usually described in terms of the Berry connection $\AAA^{(\kk)} = i \bra{\tilde{u}^{(\kk)}_{n}} \nabla^{(\kk)} \ket{\tilde{u}^{(\kk)}_{n}}$~\cite{resta2000}. Integrals (Berry phases $\vartheta_{C}$) of the Berry connection around closed paths $C$ in the Brillouin zone are invariant under changes to the phases of the Bloch states $\ket{\tilde{\psi}^{(\kk)}_{n}}$. This implies that $\bar{\rr}_n$, equal to the average value of $\AAA^{(\kk)}$ across the Brillouin zone, is also invariant~\cite{berry1984}. A further invariant quantity is given by $\BB= \mathbf{\nabla} \times \AAA^{(\kk)}$, called the Berry curvature. Local values of $\AAA^{(\kk)}$, however, depend on the phases of the Bloch states, which determine the phase-dependent part of the spread $\Omega^{n}_{D}$. It is known that this spread is minimized when the divergence of the connection vanishes, $\mathbf{\nabla} \cdot \AAA^{(\kk)} = 0$, and the minimum possible spread depends only on the Berry curvature $\BB$.

In particular, if $\BB = 0$, then the minimum possible spread $\Omega^{n}_{D}$ is zero. It follows that all Berry phases are zero and it is possible to smooth $\AAA^{(\kk)}$ such that it is uniform, at which point $\Omega^{n}_{D}=0$. To smooth the connection $\AAA^{(\kk)}$, we use a progressive phase update method: it consists of taking a succession of closed loops through the Brillouin zone, and, for each, altering the Bloch phases at points along the loop such that the projections of $\AAA^{(\kk)}$ along it are constant. This constant value is fixed by their integral around the loop, the Berry phase, which is invariant. Adjusting the phases in this way for several loops, given in \apr{app:our_algorithm2}, will result in a flattened connection, if possible.

For non-zero $\BB$, absolute uniformity of the connection $\AAA^{(\kk)}$ is not possible. However, in an attempt to suppress the divergence of the connection and therefore approach the minimum spread $\Omega^{n}_{D}$, we still choose to smooth out $\AAA^{(\kk)}$ using the progressive phase update method and find this greatly reduces $\Omega^{n}_{D}$. To achieve the minimum, we follow the progressive phase updates with the steepest-descent minimization algorithm of Marzari and Vanderbilt (cf. \secr{sec:steepest_descent}), when only terms corresponding to $\Omega^{n}_{D}$ contribute to the gradient $\Gamma^{(\kk)}$.

A particular case of interest is a system with inversion symmetry and a current gauge $U^{(\kk)}$ that is diagonal, i.e., the bands have not been mixed. As a result of the symmetry, the mode functions $\tilde{u}^{(\kk)}_{n} (\rr) \propto u^{(\kk)}_{n} (\rr) $ must be real up to a global phase, at which point the Berry curvature and thus the spread $\Omega^{n}_{D}$ vanishes. Note that optical-lattice potentials usually possess inversion symmetry since this is inherited from the lasers that created them; superlattice techniques are required to break this.

We found that even if $U^{(\kk)}$ is not diagonal, e.g., after the inter-band spread is reduced, the output of the disentangling procedure often still had zero Berry curvature for each band and the progressive phase update method reduced $\Omega^{n}_{D}$ to zero. Specifically, this occurred whenever the degeneracies in our $J$-band subspace were a result of purely geometric symmetries. We hypothesize this is a general feature, also hinted at in the results of Refs.~\cite{zak1982, brouder2007, panati2011}.

\subsection{Full procedure}\label{sec:algorithm}
Having described the elements of our computational approach, we now describe how they are pieced together. The full procedure for calculating the maximally-localized generalized Wannier states is shown in \fir{Fig:wannier-flow-diagram}.

First we calculate the band structure. Then we minimize the intra-band spread via a progressive phase update, followed by a restricted version of the steepest-descent method (for potentials with inversion symmetry, the steepest-descent part is unnecessary). At this point we are at the gauge corresponding to the maximally-localized ordinary Wannier states. We then reduce the inter-band spread using the method adapted from Souza {\em et al.}~\cite{souza2001}. The inter-band spread reduction usually has the side-effect of increasing the intra-band spread slightly, so we again apply a progressive phase update, followed by the restricted version of the steepest-descent method to minimize the intra-band spread. The above forms our initialization procedure. If this has not already found the global minimum of spread, we find that it is sufficiently close that the full steepest-descent algorithm~\cite{marzari1997} returns the maximally-localized generalized Wannier states with a close to perfect success rate.
\begin{figure}[ptb]
\begin{center}
\includegraphics[width=8.5cm]{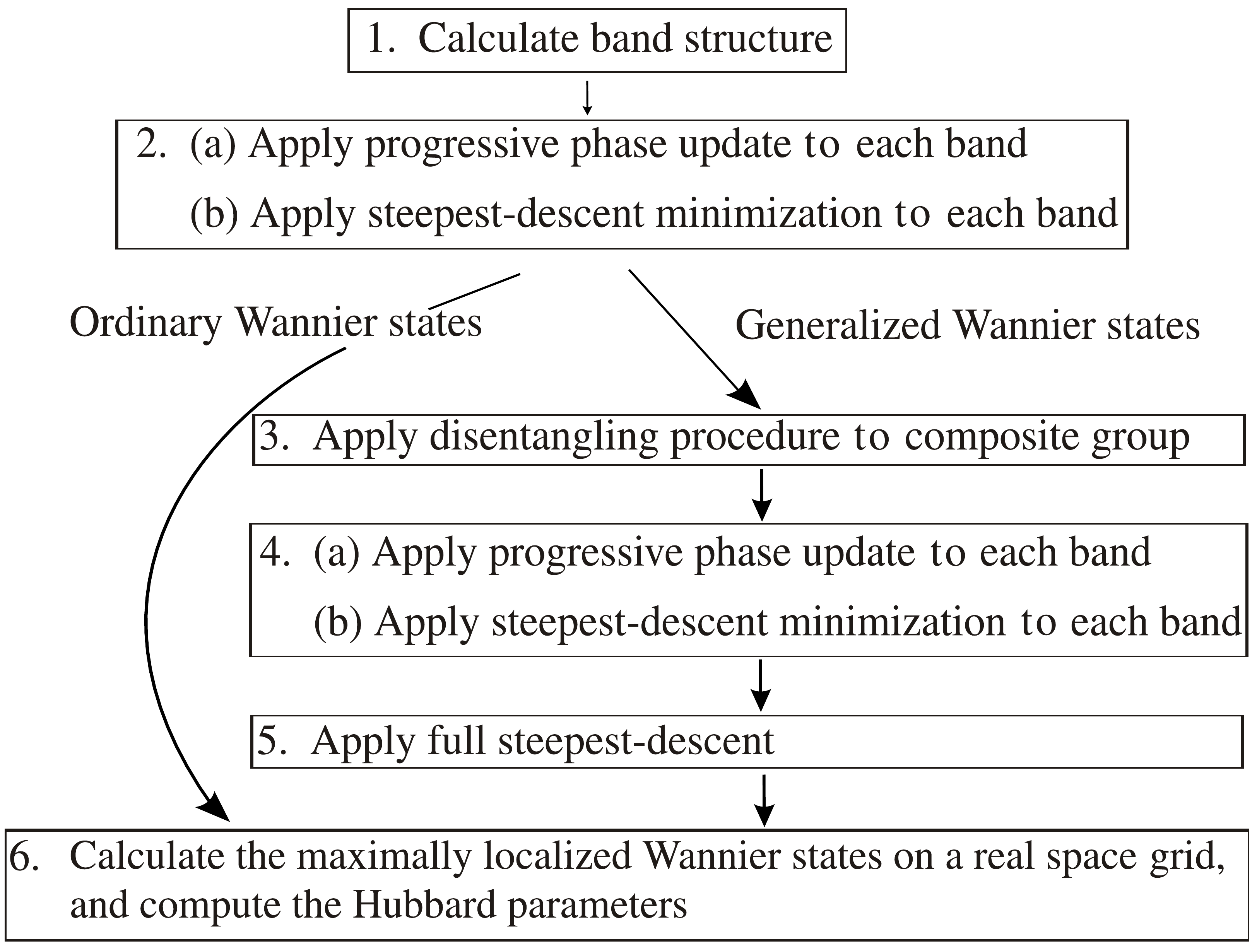}
\end{center}
\caption{Flow diagram of how our software package calculates the maximally-localized generalized Wannier states. First, it calculates the band structure, and from this computes the matrix elements $ M^{(\mathbf{k,b})}_{mn} =\sum_\GG c^{(\kk,\GG) \dagger}_{m} c^{(\kk + \bb,\GG)}_{n} $. Second, it minimizes $\Omega^{n}_{D}$ for each band as far as possible without mixing the bands. Third, it reduces $\Omega_{OD}$. Fourth, it again minimizes $\Omega^{n}_{D}$ for each band as far as possible without further mixing the bands. Fifth, it minimizes the total $\Omega$ to its global minimum via steepest-descent minimization. Last, we compute the Hubbard parameters and construct the maximally localized generalized Wannier states.}
\label{Fig:wannier-flow-diagram}
\end{figure}

\section{Results}\label{sec:results}
We next use the above procedure to derive {\em ab initio} the Hubbard Hamiltonians realized by bosons in a variety of optical-lattice potentials. The reasons are four-fold. First, we test the accuracy of our procedure. Second, we compare the procedure against others, e.g., methods using ordinary rather than generalized Wannier states. Third, we demonstrate that local Hubbard Hamiltonians can be justified for several experimentally-important cold-atom optical-lattice systems. Fourth, we provide the relevant model parameters accurately in terms of well known control parameters like the laser intensity.

For the testing, we use the hopping parameters from our derived Hubbard Hamiltonian to calculate an interpolated band-structure according to the tight-binding model. The legitimacy of our approximations can then be considered by comparing the interpolated band-structure to the original. Note that this only allows us to determine the accuracy of the hopping parameters, not the interaction terms. Since we are unable to directly verify the accuracy of discarding interaction parameters, we only discard those of magnitude equal to, or less than, that of the discarded hopping parameters for some typical range of interaction strengths $g \lesssim \tilde{g} = E_R \lambda^D$, where
\begin{equation}\label{Eq:recoil_energy}
E_\textrm{R}= \frac{h}{2 \mu \lambda^2} , 
\end{equation}
is the recoil energy, and $\lambda$ is the `averaged' wavelength of the laser beams creating the optical lattice~\footnote{The `averaged' here acknowledges that the wavelengths of the lasers used to produce the lattice must be slightly detuned from each other to avoid unwanted interference.}.

We now obtain the maximally-localized Wannier states and nearest-neighbor Hubbard models for atoms in several one- and two-dimensional optical-lattice potentials. We leave three-dimensional potentials for a future presentation.

\subsection{One-dimensional systems}\label{sec:1d_mlgwf}
\begin{figure}[ptb]
\begin{center}
\includegraphics[width=8.5cm,height=2.7cm]{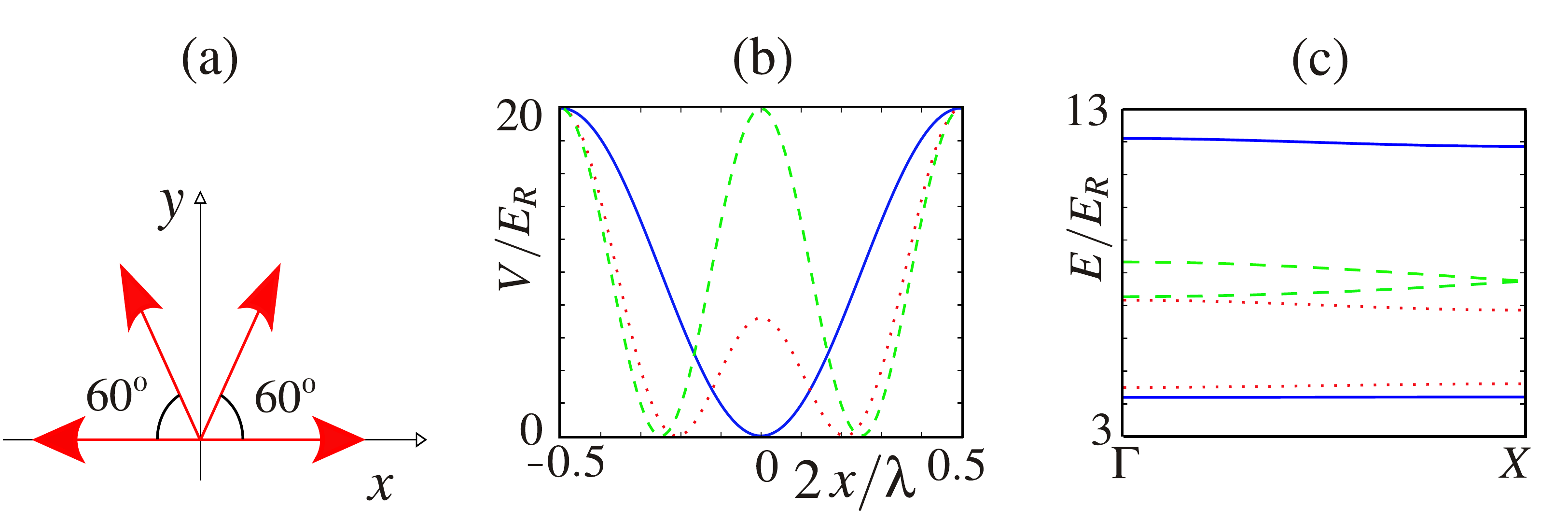}
\end{center}
\caption{(Color online) One-dimensional superlattice. (a) The configuration of lasers red-detuned from wavelength $\lambda$ to produce the superlattice potential. (b) The potential over the unit cell, for $s=0$ (blue solid line), $s=0.5$ (red dotted line) and $s=1$ (green dashed line). (c) The band-structure corresponding to the potentials in (b).}
\label{1D-super-rd-bands}
\end{figure}
\begin{figure}[ptb]
\begin{center}
\includegraphics[width=8.5cm,height=2.7cm]{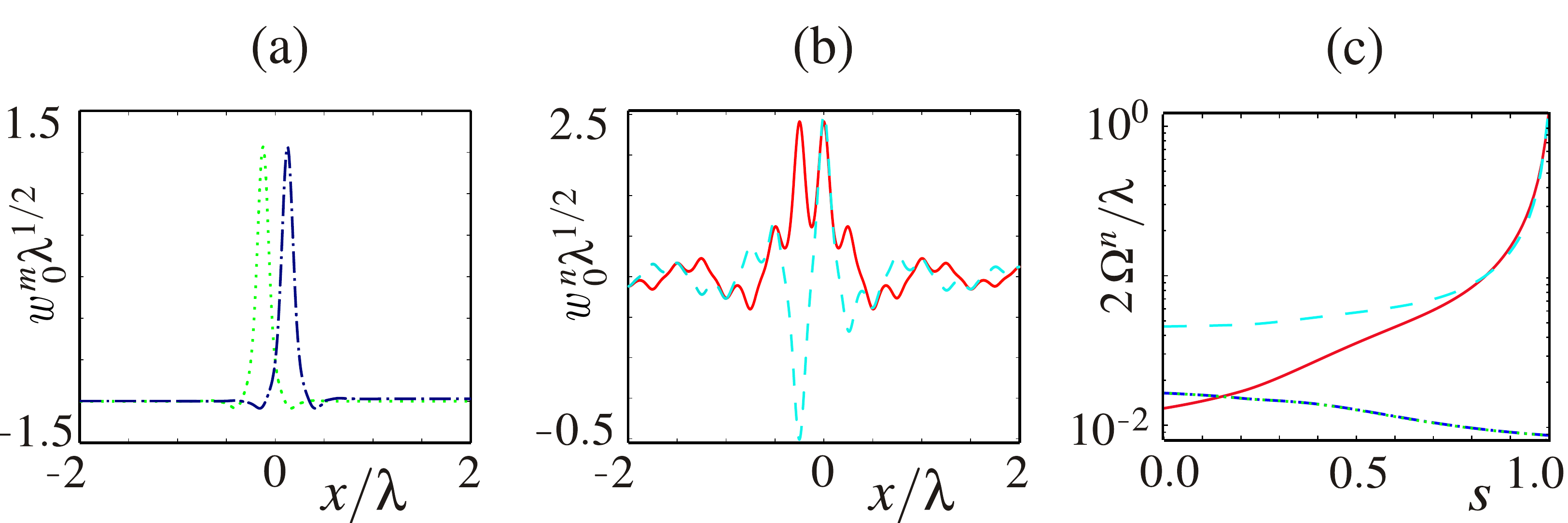}
\end{center}
\caption{(Color online) Maximally-localized Wannier states for the one-dimensional superlattice with $V_0=20 E_R$ and $s=0.999$. (a) The dotted green and dotted-dashed blue lines are the $m=1,2$ \emph{generalized} Wannier states. (b) The solid red and dashed light blue lines are the $n=1,2$ maximally localized \emph{ordinary} Wannier states. (c) The spreads of the maximally localized ordinary and generalized Wannier states as a function of $s$ (line type is the same as in (a) and (b)).}
\label{1D-super-rd-wannier}
\end{figure}
\begin{figure}[ptb]
\begin{center}
\includegraphics[width=8.5cm,height=2.7cm]{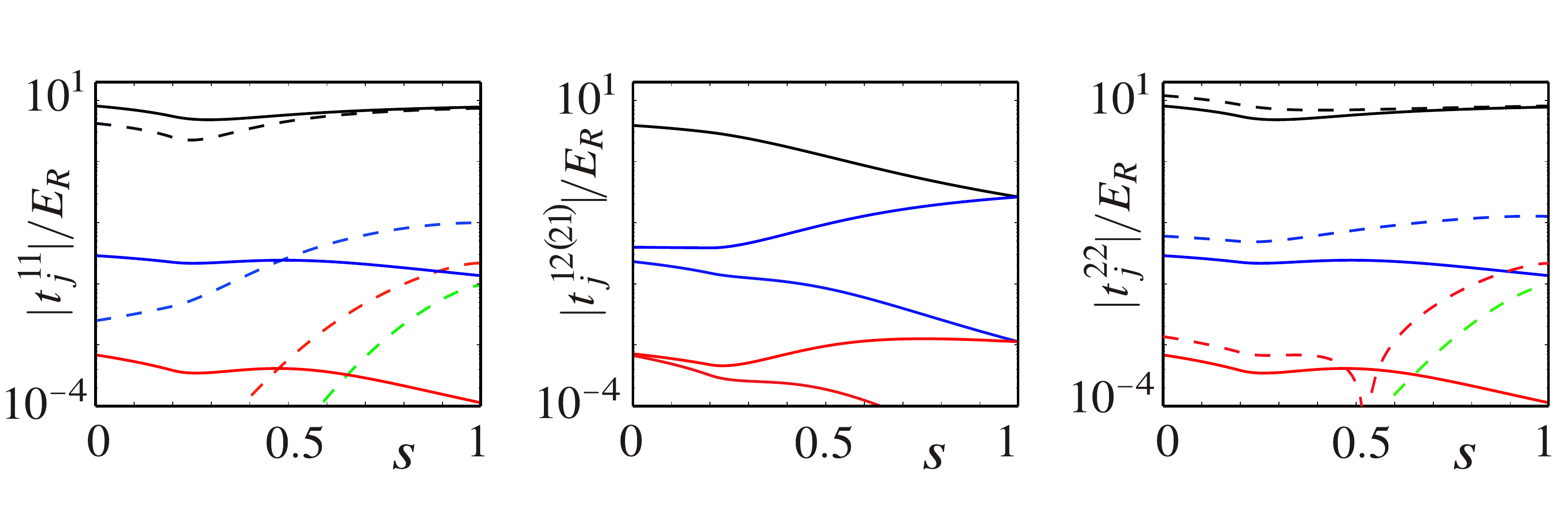}
\end{center}
\caption{(Color online) Hopping parameters for the one-dimensional superlattice. The lines show the magnitudes $|t_j^{mn}|=|t_{\0 \RR}^{mn}|$ with $2 |\RR| = j \lambda$. The black, blue, red, and green lines correspond to $j=0,1,2,3$, respectively. The solid (dashed) lines correspond to the maximally-localized generalized (ordinary) Wannier states.}
\label{1D-super-rd-hopping}
\end{figure}
\begin{figure}[pt]
\begin{center}
\includegraphics[width=8.5cm,height=2.7cm]{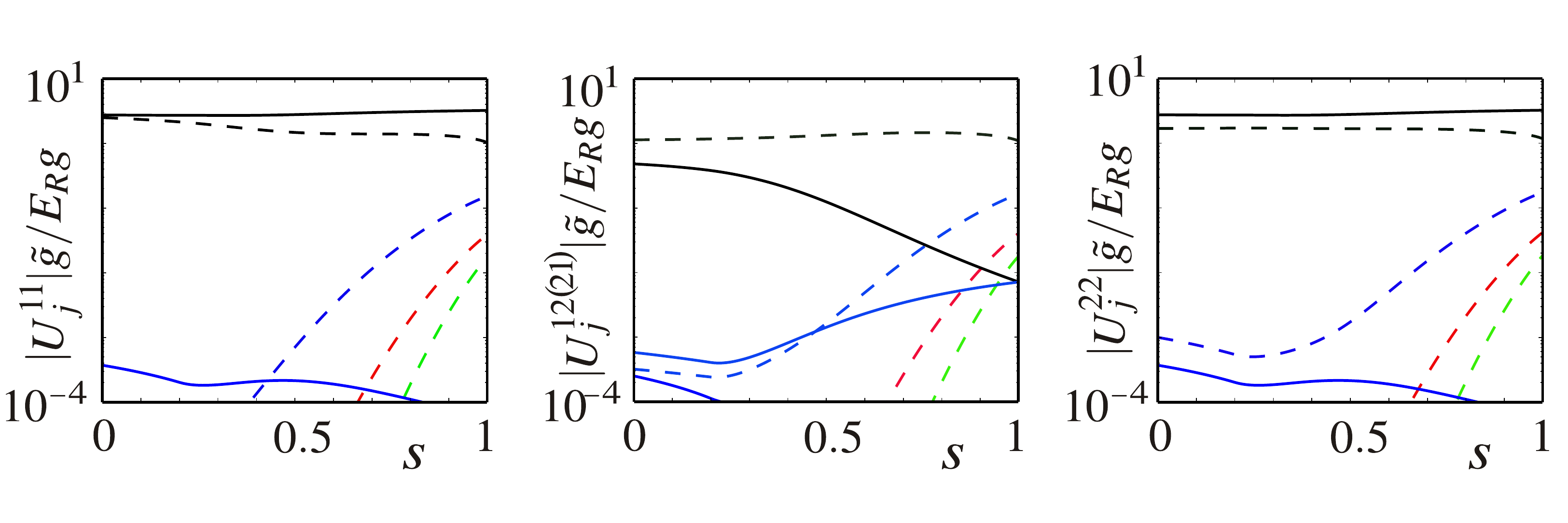}
\end{center}
\caption{(Color online) Interaction parameters for the one-dimensional superlattice. The lines show the magnitudes $|U^{mn}_{j}| = |U_{\0 \0 \RR \RR}^{mmnn}|$ of interactions between two particles in bands $m$ and $n$ at sites separated by $2 |\RR| = j \lambda$. The black, blue, red, and green lines correspond to $j=0,1,2,3$, respectively. The solid (dashed) lines correspond to the maximally-localized generalized (ordinary) Wannier states.}
\label{1D-super-rd-interaction}
\end{figure}
\begin{figure}[ptb]
\begin{center}
\includegraphics[width=8.5cm,height=2.7cm]{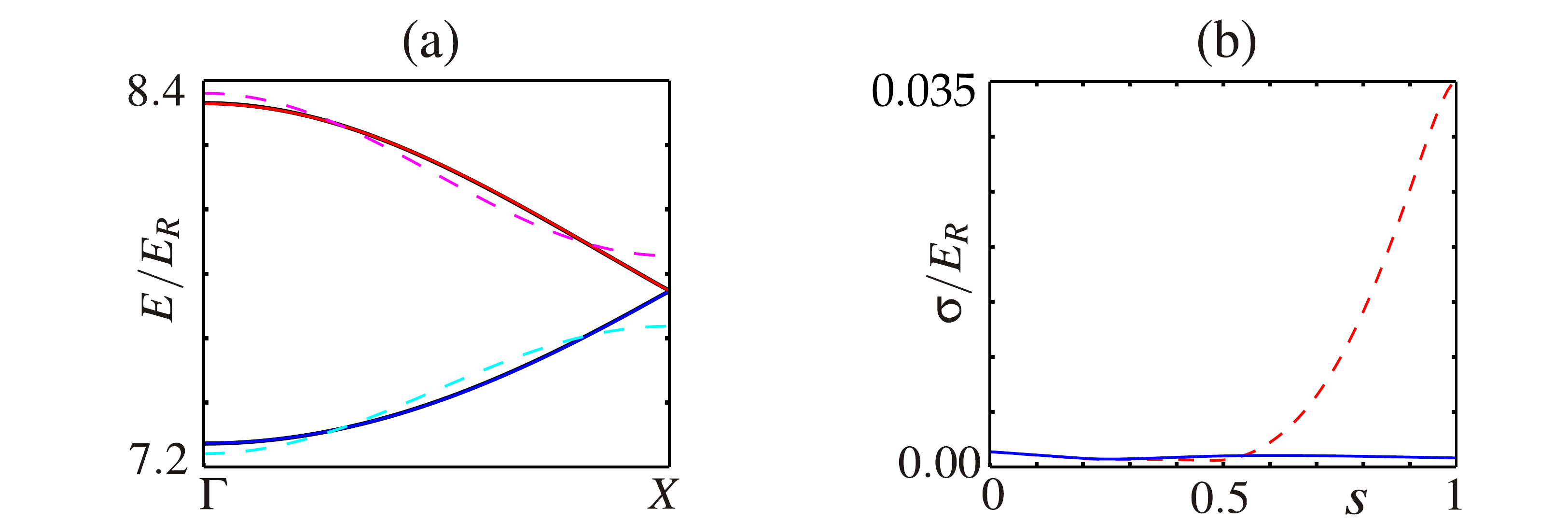}
\end{center}
\caption{(Color online) Accuracy of the one-dimensional superlattice Hubbard models. (a) Lowest and first excited bands for the superlattice potential with $V_0=20E_R$ and $s=0.999$. The blue and red solid (cyan and magenta dashed) lines show the interpolated lowest and first excited bands, respectively, for a tight-binding model using the maximally-localized generalized (ordinary) Wannier states. For maximally-localized generalized Wannier states, the interpolated bands are indistinguishable from the exact bands on this scale. (b) The standard deviation $\sigma$ between the exact bands and the interpolated bands as a function of the superlattice parameter $s$. The solid blue (dashed red) line is for maximally-localized generalized (ordinary) Wannier states.}
\label{1D-super-rd-bands-comp}
\end{figure}
To begin, we find the maximally-localized generalized Wannier states and related Hubbard parameters for bosons in a one-dimensional superlattice potential, given by
\begin{displaymath}
V(x) = V_0\left[(1-s)\sin^2(2 \pi x/ \lambda) + s\sin^2(4 \pi x/ \lambda)\right] .
\end{displaymath}
Such a potential can be produced using two independent pairs of laser beams, each red-detuned from wavelength $\lambda$ and at an angle to each other, as shown in \fir{1D-super-rd-bands}(a). Their total intensities determine the potential depth $V_0$ and their relative intensities determine the superlattice parameter $0 \le s < 1$. The lattice parameter is $\lambda / 2$. This system has been experimentally realized in Refs.~\cite{sebby2006, peil2003}, and was proposed in Ref.~\cite{vaucher2007} as a method for initialising a quantum register on a time-scale that is an order of magnitude smaller than the conventional quantum-freezing of a superfluid to a Mott insulator state~\cite{greiner2002, jaksch1998}.

The potentials for three values of $s$ are shown in \fir{1D-super-rd-bands}(b), where $V_0=20 E_\textrm{R}$. For $s = 0$ the potential is sinusoidal with a minimum at the center of each primitive cell. For $s \neq 0$ there are two minima in each primitive cell, which move either side of the center. As $s \rightarrow 1$ the potential approaches a sinusoid with lattice parameter $\lambda/4$.

The band structures for the same parameters are shown in \fir{1D-super-rd-bands}(c). For all $0 \le s < 1$ the two lowest lying bands are well separated from the higher bands, and also are not degenerate amongst themselves. Therefore, ordinary Wannier states will provide an exponentially-localized basis. We use this superlattice potential then to demonstrate that using generalized Wannier states can further localize the Wannier states even when there are no inter-band degeneracies.

We expect the benefits of generalized over ordinary Wannier states to be most notable in the $s\rightarrow 1$ limit, where the bands are close together. The maximally-localized generalized Wannier states and ordinary Wannier states for the case $s=0.999$ and $J=2$ are presented in Figs. \ref{1D-super-rd-wannier}(a) and (b), respectively. It is clear from inspection that the generalized Wannier states are more localized. This improved localization occurs for even small $s$ but is very significant for moderate or large $s \gtrsim 0.5$, as is shown in \fir{1D-super-rd-wannier}(c), which plots the respective spreads as a function of $s$.

Another way to see the effects of improved localization is to look at the magnitudes of the hopping and interaction parameters. These are shown in \fir{1D-super-rd-hopping} (hopping) and \fir{1D-super-rd-interaction} (interaction), as a function of $s$, for both the ordinary and generalized Wannier states. For large $s\gtrsim 0.5$, non-local Hubbard parameters are significantly reduced when using generalized Wannier states. This comes at the expense of allowing inter-band hopping. 

From these values it is clear that using either ordinary or generalized Wannier states, a tight-binding Hamiltonian
\begin{align}
\hat{H}_{\text{HM}} = \sum_j\sum_{n=1}^2 \bigg\{ &  - t_0^{nn} \hat{b}_j^{n \dagger} \hat{b}_j^{n}-  t_1^{nn} \hat{b}_j^{n\dagger} \left( \hat{b}_{j+1}^{n} + \hat{b}_{j-1}^{n} \right) \nonumber \\ 
&+ \tfrac{1}{2}U_{0}^{nn} \hat{b}_j^{n \dagger} \hat{b}_j^{n \dagger} \hat{b}_j^{n} \hat{b}_j^{n}  + \sum_{m=1,m \neq n}^{2} [ - t_0^{mn} \hat{b}_j^{m \dagger} \hat{b}_j^{n} \nonumber \\
&- t_1^{mn}\hat{b}_j^{m\dagger}\left(\hat{b}_{j+1}^{n} + \hat{b}_{j-1}^{n} \right) + \tfrac{1}{2}U_{0}^{mn} \hat{b}_j^{m \dagger} \hat{b}_j^{m} \hat{b}_j^{n \dagger} \hat{b}_j^{n}] \bigg\} \nonumber,
\end{align}
can be derived and justified from first principles. Here, for clarity, we have replaced the label $\RR$ by the label $j = 2 |\RR| / \lambda$. The model derived using generalized Wannier states is more accurate, as we can demonstrate by comparing interpolated bands to the original. In \fir{1D-super-rd-bands-comp}(a) this is shown for both maximally-localized ordinary and generalized Wannier states, and superlattice parameter $s=0.999$. The generalized Wannier states almost exactly reproduce the band structure, while there are significant deviations for the Hamiltonian derived using ordinary Wannier states. In \fir{1D-super-rd-bands-comp}(b) we show the standard deviation,i.e., the average root mean squared error of the energies averaged over the bands, between the interpolated bands and exact bands as a function of $s$. This demonstrates that the difference in accuracy between using ordinary and generalized Wannier states is appreciable for $s \gtrsim 0.5$. In fact, we should have expected this from the non-sinusoidal nature of the bands for large $s$. A tight-binding model built from ordinary Wannier states can only ever result in a sinusoidal band structure. Generalized Wannier states and inter-band hopping they describe have no such restriction.

These results confirm that for $s \gtrsim 0.5$, the accuracy of the local model found using generalized Wannier states becomes significantly better than that using ordinary Wannier states. The reason for this difference is that the two generalized Wannier states can each break the reflection symmetry in the primitive cell to localize around a different minimum (see Fig.~\fir{1D-super-rd-wannier}(a)). Meanwhile the maximally-localized ordinary Wannier states cannot break this symmetry and instead are symmetric and antisymmetric combinations of two functions localized at each of the minima (see Fig.~\fir{1D-super-rd-wannier}(b)). Aside from leading to more accurate local Hubbard models, the use of generalized Wannier states are more relevant for cold atom experiments. In such experiments, it is the presence of a particle at a position in space rather than the symmetry of its wavefunction that is measured through high-resolution imaging~\cite{bakr2009}. Thus a Hubbard model corresponding to atoms in spatially-separated sites is preferable to atoms in symmetric/antisymmetric superpositions.  We similarly expect generalized Wannier states to be important for other lattices that possess more than one potential minimum per primitive cell.

\subsection{Two-dimensional systems}\label{sec:mlgwf2d}
The use of generalized Wannier states is paramount in two dimensions, as degeneracies in the lowest bands are likely to occur as a result of crystallographic point-group symmetries. Hence the maximally-localized ordinary Wannier states could fail to provide an exponentially localized basis due to the resulting non-analyticity of the bands. Further, we will see cases where the maximally-localized generalized Wannier states are not centered around inversion points of the lattice, and do not share the symmetry of the lattice. In these cases, approximating the states using Gaussian functions would lead to a particularly inaccurate estimate of the Hubbard parameters.

To showcase our procedure we now calculate accurately and from first principles the maximally-localized generalized Wannier states and Hubbard parameters for atoms in an optical lattice, with either hexagonal or Kagom\'{e} geometries. Both potentials have multiple minima per primitive cell and lead to a degenerate set of lowest bands, thus representing a significant challenge using any other method. Both of these structures also play an important role in condensed-matter physics, see e.g. Refs.~\cite{panahi2011, tarruell2012, santos2004, ruostekoski2009, jo2012}.

\subsubsection{Hexagonal lattice}\label{Sec:2D-hexagonal-lattice}
\begin{figure}[ptb]
\begin{center}
\includegraphics[width=8.5cm,height=5.4cm]{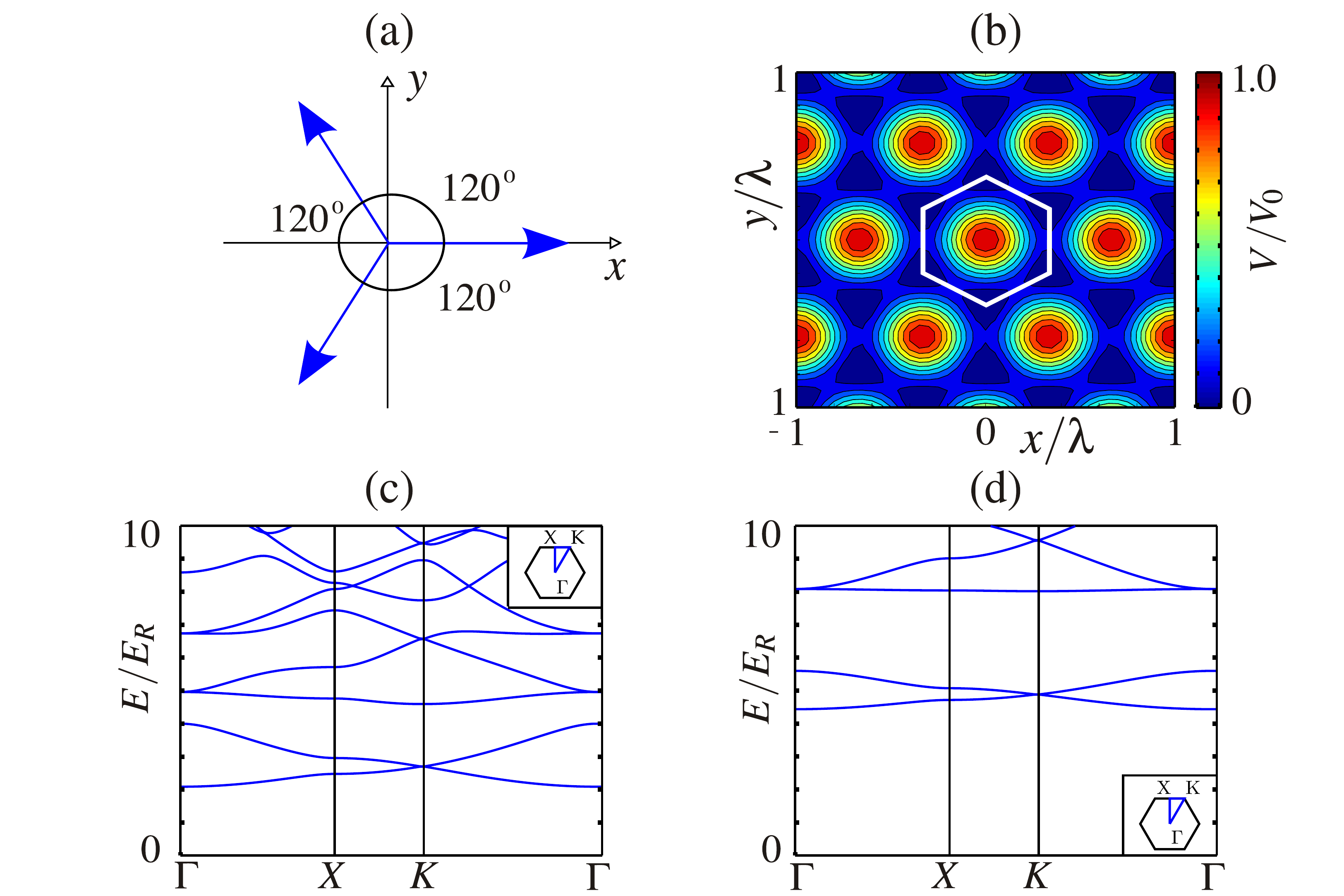}
\end{center}
\caption{Hexagonal lattice. (a) The beam configuration for generating the optical lattice. The three beams are blue-detuned from the wavelength $\lambda$. (b) The lattice potential, with the white line marking the boundary of the Wigner-Seitz unit cell. (c) The band-structure for lattice depth $V_0=10E_\textrm{R}$. The energies are displayed along the path through the Brillouin zone shown in the inset. (d) Similarly for $V_0= 30E_\textrm{R}$.}
\label{Fig:2D-hexagonal-potential}
\end{figure}
\begin{figure}[ptb]
\begin{center}
\includegraphics[width=8.5cm,height=2.7cm]{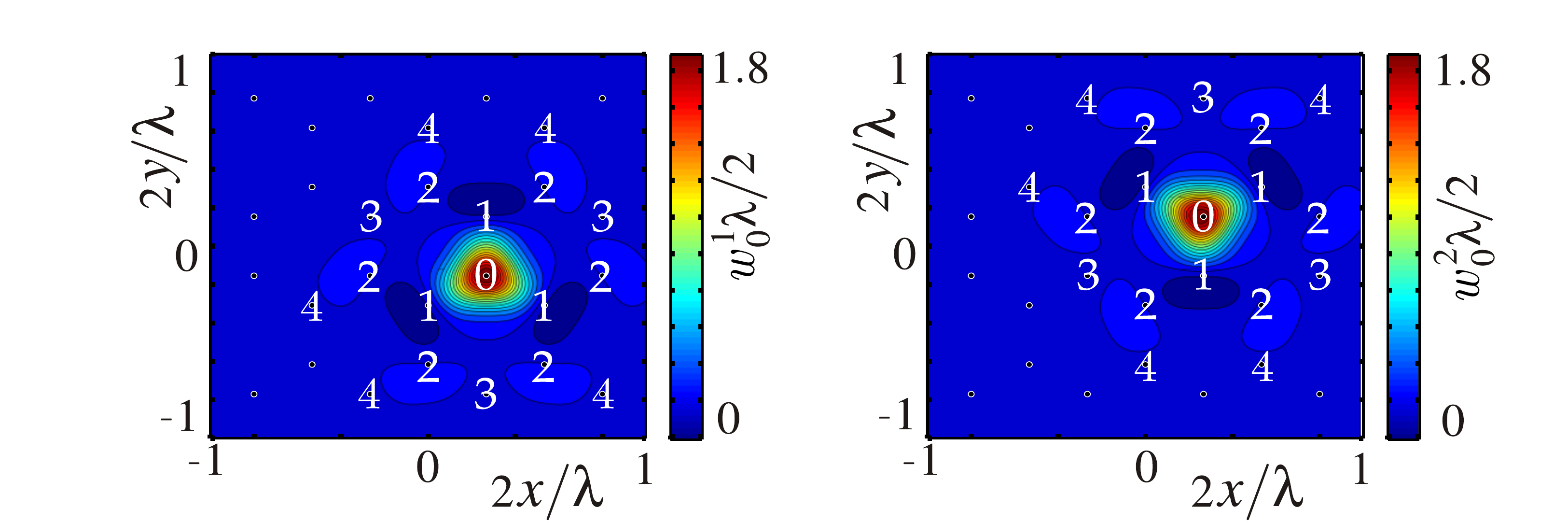}
\end{center}
\caption{Maximally-localized generalized Wannier states for the hexagonal lattice. The two lowest bands are shown for lattice depth $V_0=10E_\textrm{R}$. We have labeled the potential minima with equal hopping and interaction parameters from the `home' minimum by $j=0,1,2,3,4$.}
\label{Fig:2D-hexagonal-wannier}
\end{figure}
\begin{figure}[ptb]
\begin{center}
\includegraphics[width=8.5cm,height=2.7cm]{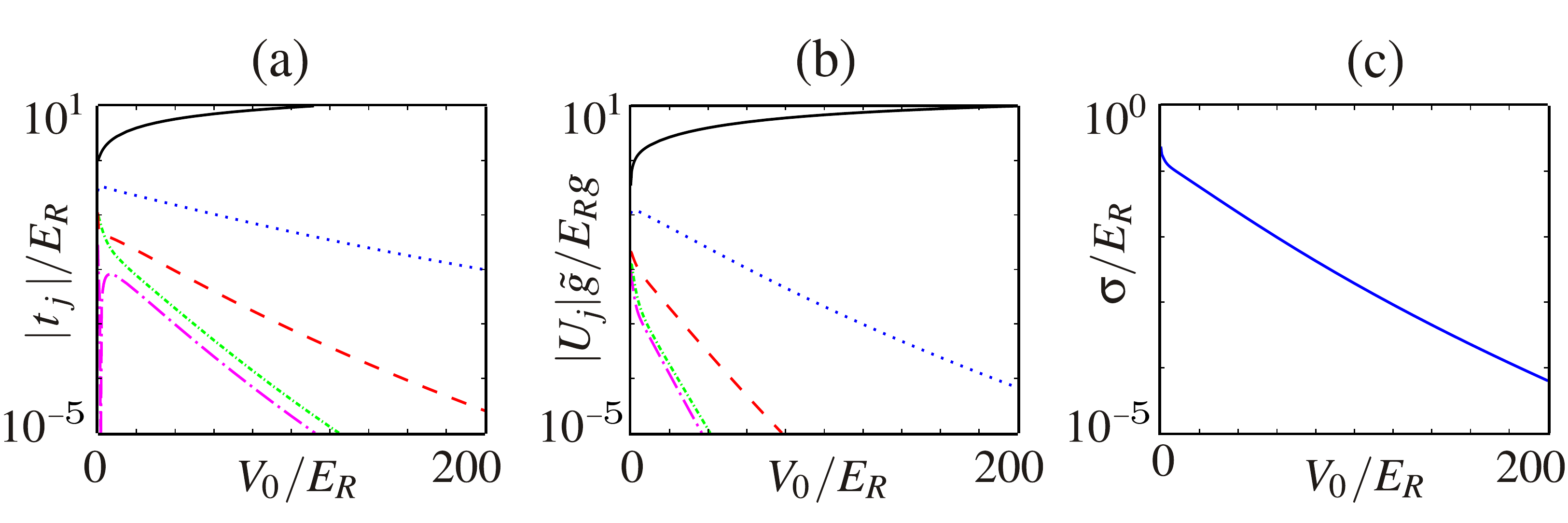}
\end{center}
\caption{(Color online) Hopping and interaction parameters for the hexagonal lattice. (a) The magnitudes $|t_j| = |t_{\0 \RR}^{mn}|$ of the hopping parameters, as a function of lattice depth $V_0$, where the centers of $\ket{\0 m}$ and $\ket{\RR n}$ are the $j$-th smallest distance from each other (cf. \fir{Fig:2D-hexagonal-wannier}). The black solid, blue dotted, red dashed, green dot-dashed, and magenta dot-long dashed lines are for $j=0,1,2,3,4$ respectively. (b) Similarly for the magnitudes $|U_j| = |U_{\0\0 \RR \RR}^{mmnn}|$ of the interaction parameters. (c) The total standard deviation $\sigma$ between the exact lowest bands and the interpolated tight-binding bands as a function of lattice depth.}
\label{Fig:2D-hexagonal-elements}
\end{figure}

Three blue-detuned beams of approximately equal wavelength $\lambda$, shown in \fir{Fig:2D-hexagonal-potential}(a), generate a hexagonal optical-lattice potential, written as
\begin{align*}
V (x,y) = \frac{V_0}{9}\big[  & 3 + 2\cos\bigg(\frac{2 \sqrt{3} \pi y}{\lambda}\bigg) \nonumber \\ &+ 4\cos\bigg(\frac{3 \pi x}{ \lambda}\bigg)\cos\bigg(\frac{\sqrt{3} \pi y}{\lambda}\bigg)\big],
\end{align*}
and plotted in \fir{Fig:2D-hexagonal-potential}(b). The potential exhibits two minima per unit cell, positioned at cell vertices that form a hexagonal (honeycomb) structure. The consequence of there being two potential minima per unit cell is that the two lowest bands are degenerate at the K points of the Brillouin zone, as shown in Figs.\ \ref{Fig:2D-hexagonal-potential}(c) and (d) for lattice depths $V_0=10E_\textrm{R}$ and $V_0=30E_\textrm{R}$, respectively.

The maximally-localized generalized Wannier states for the two lowest bands, using $J=2$, are shown in \fir{Fig:2D-hexagonal-wannier} for a lattice depth $V_0=10E_\textrm{R}$. Both states possess three-fold rotational symmetry about their centers and are images of one another through a rotation of $60^\circ$ about the center of the Wigner-Seitz unit cell. As for the one-dimensional superlattice, both generalized Wannier states are localized around a potential minimum, rather than at the Wyckoff positions (centers of inversion). As a result $\Omega_\textrm{D}\neq0$ for the pair although the total spread is minimized and the cell-periodic superposed states (cf. \eqr{eq:periodic_superpositions}) are real. Since inversion symmetry is broken it is clear that a Gaussian function would not adequately describe these Wannier states even in the deep lattice limit.

The magnitudes of the hopping and interaction parameters for the two lowest bands are shown in \fir{Fig:2D-hexagonal-elements}. Since the maximally-localized generalized Wannier states are related through a symmetry operation the parameters within each band are identical. We therefore label the parameters not by site and band, but by $j$, the rank of the distance between potential minima, as shown in \fir{Fig:2D-hexagonal-wannier}. Parameters for $j=0,2$ are intra-band, while those for $j=1,3,4$ are inter-band. We calculate these parameters up to a lattice depth of $V_0=200E_\textrm{R}$. We observe that for large $V_0$ the significant parameters are the on-site interaction parameter $U_{0}$ and the hopping parameter $t_1$. The interaction parameter $U_{1}$ corresponding to the interaction of Wannier states in neighboring potential minima is also relatively large but is at least an order of magnitude less than $t_1$ for $V_0 \gtrsim 10E_\textrm{R}$ and typical interaction strengths $g \approx \tilde{g}$. With these observations, the Hamiltonian for the hexagonal optical lattice is accurately represented by
\begin{equation}
\hat{H}_{\text{HM}} = - \sum_i t_0 \hat{b}_i^\dagger \hat{b}_i - \sum_{\langle i,j\rangle} t_1 \hat{b}_i^\dagger \hat{b}_j + \sum_i\tfrac{1}{2}U_{0} \hat{b}_i^\dagger \hat{b}_i^\dagger \hat{b}_i \hat{b}_i,
\label{2D-hexagonal-Hamiltonian-gen}
\end{equation}
where the sums are taken over potential minima (we have now dropped the band index and instead labeled the minima by the indices $i$ and $j$), each with three nearest-neighbors, which we denote by the angled brackets.

We once again insert the hopping parameters included in the Hamiltonian into a tight-binding model to recreate the single-particle band-structure. The standard deviation between the interpolated bands and the exact bands is shown in \fir{Fig:2D-hexagonal-elements}(c) as a function of lattice depth. This again decreases exponentially with lattice depth indicating the high accuracy of the model at all but shallow depths.

\subsubsection{Kagom\'{e} lattice}\label{Sec:2D-kagome-lattice}
\begin{figure}[ptb]
\begin{center}
\includegraphics[width=8.5cm,height=2.7cm]{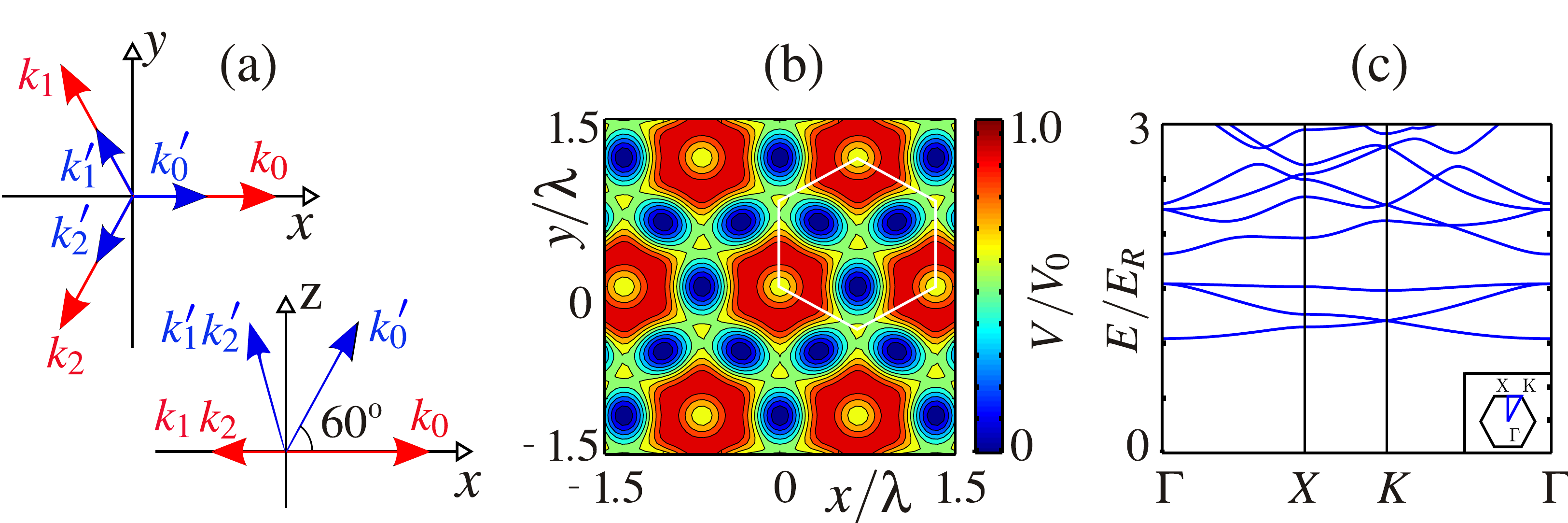}
\end{center}
\caption{(Color online) The Kagom\'{e} lattice. (a) The beam configuration for generating the optical-lattice potential. The projections of the beam wave-vectors are shown in the $x$-$y$ and $x$-$z$ planes. The beams are both red- and blue-detuned from the same wavelength $\lambda$. (b) The resulting lattice potential, with the white line marking the boundary of the Wigner-Seitz unit cell. (c) The band-structure for lattice depth $V_0=2E_\textrm{R}$. The energies are displayed along the path through the Brillouin zone shown in the inset.}
\label{Fig:2D-kagome-potential}
\end{figure}
\begin{figure}[ptb]
\begin{center}
\includegraphics[width=8.5cm,height=2.7cm]{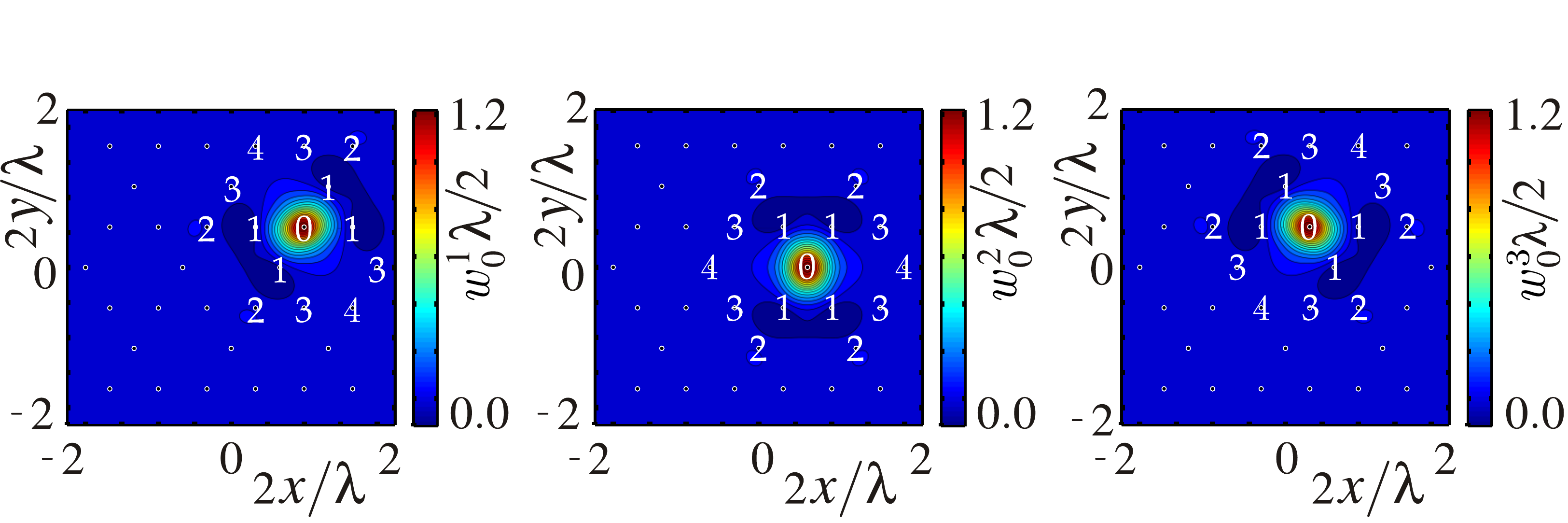}\hspace{0.3cm}
\end{center}
\caption{(Color online) Maximally-localized generalized Wannier states for the Kagom\'{e} lattice. The three lowest bands are shown for lattice depth $V_0=10E_\textrm{R}$. We have labeled the potential minima with equal hopping parameters from the `home' minimum by $j=0,1,2,3,4$. }
\label{Fig:2D-kagome-wannier}
\end{figure}
\begin{figure}[ptb]
\begin{center}
\includegraphics[width=8.5cm,height=2.7cm]{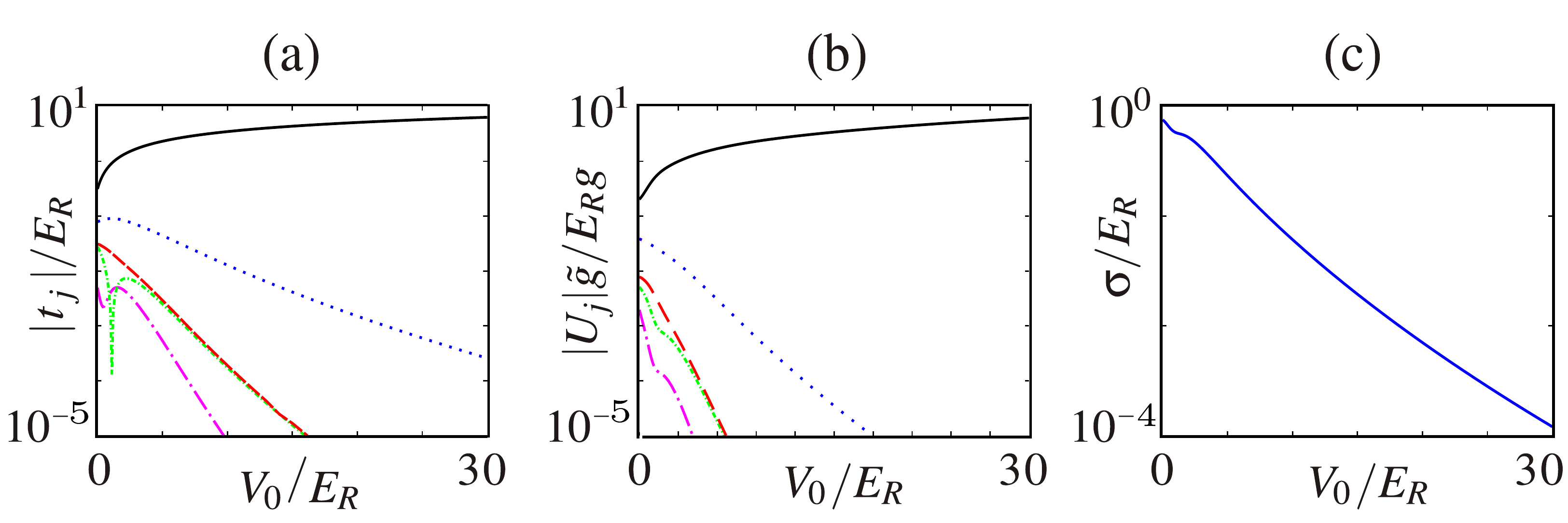}
\end{center}
\caption{(Color online) Hopping and density-density interaction parameters for the Kagom\'{e} lattice. (a) The magnitudes $|t_j| = |t_{\0 \RR}^{mn}|$ of the hopping parameters, as a function of lattice depth $V_0$, where the centers of $\ket{\0 m}$ and $\ket{\RR n}$ are the $j$-th smallest distance from each other (cf. \fir{Fig:2D-kagome-wannier}). The black solid, blue dotted, red dashed, green dot-dashed, and magenta dot-long dashed lines are for $j=0,1,2,3,4$ respectively. (b) Similarly for the magnitudes $|U_j| = |U_{\0\0 \RR \RR}^{mmnn}|$ of the interaction parameters. (c) The total standard deviation $\sigma$ between the exact lowest bands and the interpolated tight-binding bands as a function of lattice depth.}
\label{Fig:2D-kagome-elements}
\end{figure}

The Kagom\'{e} lattice has received a large degree of interest in recent years because it leads to a highly-frustrated many-body Hamiltonian~\cite{santos2004, ruostekoski2009, damski2005, jo2012}. This lattice may be created using six lasers of approximate wavelength $\lambda$, three of which are red-detuned and three of which are blue-detuned. The setup is shown schematically in \fir{Fig:2D-kagome-potential}(a) and the resulting Kagom\'{e} potential
\begin{align*}
V (x,y) \propto & -\cos\bigg(\frac{2 \sqrt{3} \pi y}{\lambda}\bigg) - 2\cos\bigg(\frac{3 \pi x}{\lambda}\bigg)\cos\bigg(\frac{\sqrt{3} \pi y}{\lambda}\bigg) \\
&+ \cos\bigg(\frac{\sqrt{3} \pi y}{\lambda}\bigg) + 2\cos\bigg(\frac{3 \pi x}{2\lambda}\bigg)\cos\bigg(\frac{\sqrt{3} \pi y}{2 \lambda}\bigg),
\end{align*}
is shown in \fir{Fig:2D-kagome-potential}(b). We scale the potential such that the full lattice depth $V_{0}$ is the difference between the maximum and minimum of the potential $V(x,y)$. The primitive unit cell possesses three potential minima, and the lowest three bands, shown in \fir{Fig:2D-kagome-potential}(c) for a lattice depth $V_0=2E_\textrm{R}$, are degenerate; two of the bands are degenerate at the K points and are reminiscent of the lowest bands of the hexagonal lattice, while the highest energy band is almost flat and is degenerate at the $\Gamma$ point.

The three maximally-localized generalized Wannier states for the three lowest bands, using $J=3$, are plotted in \fir{Fig:2D-kagome-wannier} for a lattice depth $V_0=10E_\textrm{R}$, and each is once again located at a potential minimum. Since the potential minima are located at Wyckoff positions, the generalized Wannier states possess inversion symmetry and $\Omega_\textrm{D}=0$. Each state is only two-fold symmetric under rotation in accordance with the point-symmetry of the Wyckoff position it is centered on.

The states are images of each other through a rotation of $120^\circ$ about the center of a trimer. Because of this the magnitudes of the hopping parameters, plotted in \fir{Fig:2D-kagome-elements}(a), between equivalent neighboring potential minima are equal, as was observed with the hexagonal lattice. Also similar to the hexagonal lattice, the parameters corresponding to hopping between adjacent minima decay almost exponentially, while the on-site interaction parameter dominates (see \fir{Fig:2D-kagome-elements}(b)). The nearest-neighbor interaction parameters are at least an order of magnitude smaller except at very low lattice depths. Once again, due to symmetry this leads us, for a sufficiently deep lattice, to the Hamiltonian given in \eqr{2D-hexagonal-Hamiltonian-gen}, where instead there are four nearest neighbors. Once again we can reassure ourselves of the accuracy of the derived Hamiltonian by looking at the standard deviation, shown in \fir{Fig:2D-kagome-elements}(c), between the interpolated bands and the exact bands. This decreases exponentially with lattice depth and is significantly smaller than the exact band-width, indicating good accuracy.

\section{Conclusions}\label{sec:conclusions}
We have calculated, from first principles, the parameters of nearest-neighbour Hubbard models for several optical lattice potentials, including the honeycomb and Kagom\'{e} potentials, demonstrating quantitatively for which lattice depths such models are accurate. Strongly-correlated phenomena probed in optical lattice experiments and quantum simulations depend delicately on the ratios of kinetic and interaction energies. Therefore precisely determining them {\em ab initio}, as done here, is essential for diagnosing and interpreting such experimental results and for using optical lattices as quantum simulators.

To perform our calculations we have developed a freely available software package~\cite{tnt} that, given an optical lattice potential, will efficiently calculate the corresponding maximally-localized generalized Wannier states without any prior-knowledge of their form in any spatial dimension. This will allow cold-atom researchers to easily and accurately determine Hubbard models realized by any laser setup. We hope that this tool will be useful for the optical-lattice community.

\begin{acknowledgments}
We thank the CCPQ initiative for hosting the code on the CCPForge site and the Oxford Martin School for support through the Programme on Bio-Inspired Quantum Technologies. SRC and DJ thank the National Research Foundation and the Ministry of Education of Singapore for support. This work was completed within the EuroQUAM (EP/E04162/1) project.
\end{acknowledgments}

\appendix

\section{Accuracy of the truncated Fourier representation and discretized mesh}
\label{sec:discretization}
The cut-off wave-vector corresponds to a maximum kinetic energy for the plane-wave components given by $E_\textrm{cut-off}=G_\textrm{max}^2/2 \MM$, and introduces a minimum spatial resolution $\lambda_\textrm{min}=2 \pi / G_\textrm{max}$ for describing real space functions in the system, namely, the potential, the Bloch states and the Wannier states. The minimum spatial resolution must be smaller than the spatial variations in these states in order for them to be accurately recreated using the truncated set of coefficients, therefore the cut-off energy must be at least as large as the highest energy Fourier component of the potential. One can then increase the cut-off energy until the energies for each band under consideration have converged, at which point all coefficients $c^{(\kk, \GG)}_{n}$ of significant magnitude describing the Bloch periodic functions $u^{(\kk)}_{n}(\textbf{r})$ are included. Typically, this requires the cut-off energy to be an order of magnitude greater than the upper-end of the energy range of interest. In our calculations a cut-off energy of $E_\textrm{cut-off}= 50 E_R$ is sufficient for band convergence and suitably limits the total number of coefficients such that even in three dimensions our procedure is not computationally expensive. Here $E_{R}$ is the recoil energy, defined in \eqr{Eq:recoil_energy}.

The discretization of vectors in reciprocal space corresponds to considering a finite real space lattice with periodic boundary conditions and $\texttt{M}$ primitive unit cells in each lattice direction. So we expect it to be valid when $\texttt{M}$ is large and surface effects are negligible.

\section{Algorithm for reducing inter-band spread}\label{app:our_algorithm}
\begin{figure}[ptb]
\begin{center}
\includegraphics[width=8.5cm]{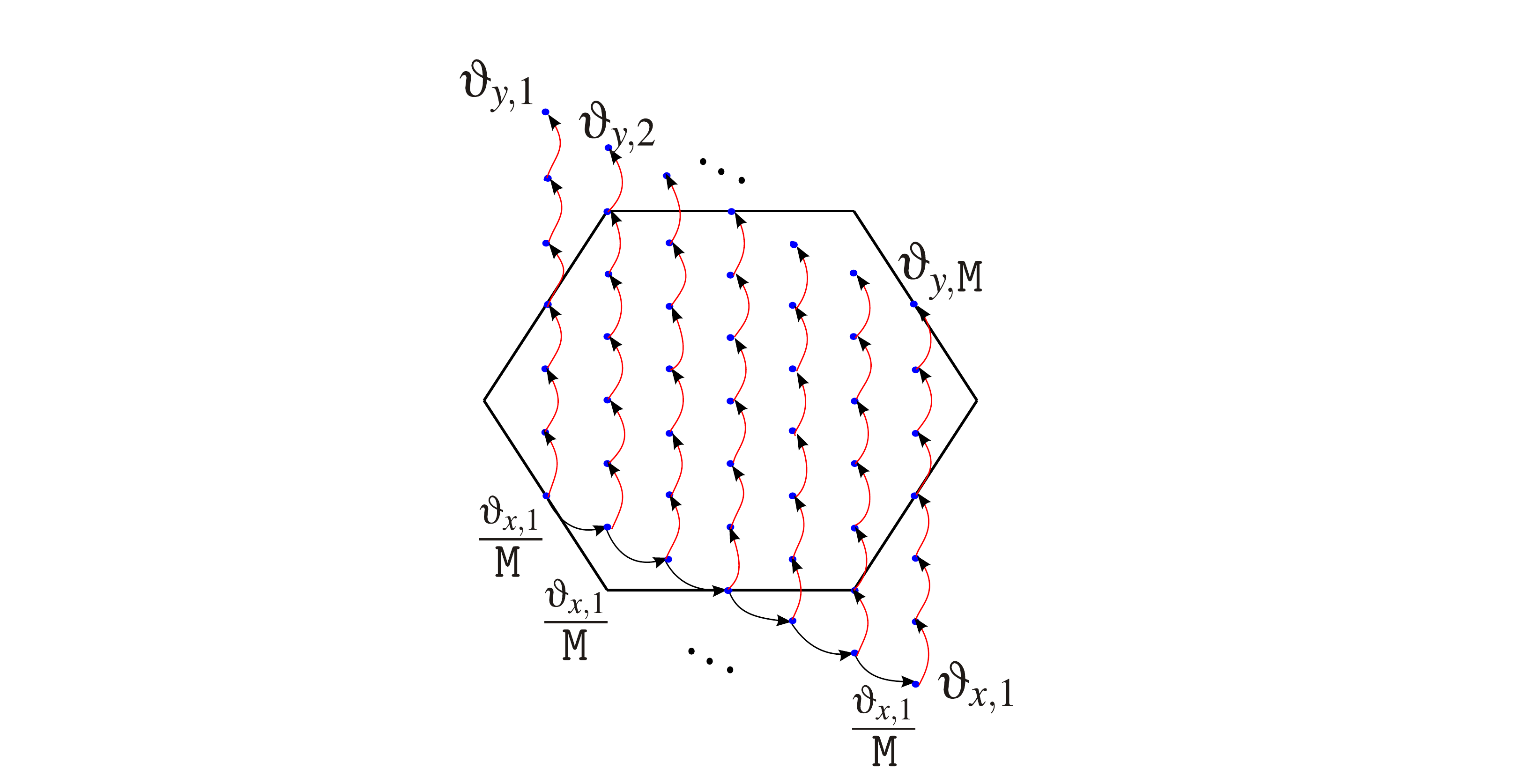}
\caption{(Color online) Progressive phase update method. Starting from the bottom left corner of the Brillouin zone mesh (blue dots), the phase of the neighbor to the right (connected by the black arrow) is adjusted such that $\textrm{Im}[\textrm{ln}M^{(\mathbf{k},\mathbf{b})}]=\vartheta_{x,1}/\MM$ for the pair, where $\vartheta_{x,1}$ is the Berry phase in this direction. The same adjustment is made for the next neighbor and so on until the end of the mesh is reached (lower-right corner). One then has $\textrm{Im}[\textrm{ln}M^{(\mathbf{k},\mathbf{b})}]=\vartheta_{x,1}/\MM$ for all mesh points along this path. The process is repeated for each path in the next reciprocal lattice direction as shown by the red arrows.}
\label{Fig:q-mesh2}
\end{center}
\end{figure}
Our method for reducing the inter-band spread involves taking, for each $n=1,\dots,J-1$ in turn, the $J-n+1$ bands $n,\dots,J$ and constructing from them an $n$-th band that is optimally smooth in $\kk$-space, such that the most localized Wannier state possible may be constructed for this band.

For each $n$, the algorithm, based on Ref.~\cite{souza2001}, proceeds as follows. We calculate the Hermitian matrices
\begin{equation}\label{eq:our_z}
Z_{mp}^{(\mathbf{k})} = \sum_{\mathbf{b}} \omega_{\mathbf{b}} M^{(\mathbf{k,b})}_{mn} M_{pn}^{(\mathbf{k,b})\ast},
\end{equation}
where $m,p$ run over $n,\ldots,J$. We then apply, for every $\kk$ in turn, a transformation $V^{(\mathbf{k})}=\mathbbm{1}_{n-1}\otimes X^{(\mathbf{k})}$, where the unitary $X^{(\mathbf{k})}$ diagonalises $Z^{(\mathbf{k})}$, i.e., $Z^{(\mathbf{k})}=X^{(\mathbf{k})}\Lambda^{(\mathbf{k})}X^{(\mathbf{k})\dagger}$, with $\Lambda^{(\mathbf{k})}$ diagonal, reducing the spread to
\begin{displaymath}
\Omega^{n}_{I,D} = \sum_{\mathbf{b}} \omega_{\mathbf{b}} - \frac{1}{M^{D}} \sum_{\mathbf{k}} \Lambda_{11}^{(\mathbf{k})}.
\end{displaymath}
The $X^{(\mathbf{k})}$ are always chosen at each $\mathbf{k}$ such that $\Lambda_{11}^{(\mathbf{k})}$ is the largest eigenvalue of $Z_{(\mathbf{k})}$, so this spread is as small as possible. The procedure in this paragraph is then applied repeatedly until convergence is achieved. The gauge for which $\Omega_{I,D}^{n}$ is minimized is a convergence point~\cite{souza2001}.

On occasion the above procedure can become unstable, and we prevent this by replacing \eqr{eq:our_z} by an equal weighting of $Z_{mp}^{(\mathbf{k})}$ calculated during the current and previous iteration. This has no effect on the locations at which the algorithm can converge.

\section{Algorithm for reducing intra-band spread}\label{app:our_algorithm2}
For the progressive phase update method, we smooth the Berry connection over loops consisting of straight lines through the Brillouin zone, in the directions of the reciprocal lattice vectors. In the reciprocal mesh representation, the Berry connection at each $\kk$ is given by $- \sum_{\mathbf{b}} \omega_{b} \mathbf{b} \text{Im} [\ln M^{(\mathbf{k,b})}_{nn}]$ and so uniformity across a straight loop $C^{(\kk',\bb)}$ going through $\kk'$ in direction $\bb$ is achieved by choosing phases such that the projection $\text{Im} [\ln M^{(\mathbf{k,b})}_{nn}]$ of the connection onto this line is the same at each point. Specifically, since integrating over the loop must give the Berry phase $\vartheta_{C^{(\kk',\bb)}} = \sum_{\kk \in C^{(\kk',\bb)} } -\text{Im} [\ln M^{(\mathbf{k,b})}_{nn}]$, we set $\text{Im} [\ln M^{(\mathbf{k,b})}_{nn}] = \vartheta_{C^{(\kk',\bb)}}/ \MM$ at each point $\kk$ on the loop. The loops and the order in which we smooth the Berry connection across them is depicted in \fir{Fig:q-mesh2}.

\end{document}